\documentclass[usenatbib,useAMS,letterpaper]{mnras}
\usepackage{float}
\usepackage{graphicx}
\usepackage{epstopdf}
\usepackage{ textcomp }
\usepackage{multirow}
\usepackage{ctable}
\usepackage{url}
\usepackage{times}
\usepackage{amsmath}
\usepackage{amssymb}
\usepackage{xspace}
\usepackage{mathrsfs} 
\usepackage{tabularx}
\usepackage{fixltx2e}
\usepackage{color}
\usepackage{placeins}
\usepackage{comment}

\newcommand{\acknowledgments}{\begin{small}\section*{Acknowledgements}\end{small}}

\defcitealias{2019MNRAS.487.4393S}{Paper {\small I}}
\defcitealias{2020MNRAS.491.1190S}{Paper {\small II}}
\defcitealias{2021MNRAS.507..175S}{Paper {\small III}}
\newcommand\sref[1]{\hyperref[#1]{Section~\ref*{#1}}}
\newcommand\fref[1]{\hyperref[#1]{Fig.~\ref*{#1}}}
\newcommand\Eqref[1]{Eq.~(\hyperref[#1]{\ref*{#1}})}
\newcommand\eeqref[1]{Eq.~\hyperref[#1]{\ref*{#1}}}

\newcommand\tref[1]{\hyperref[#1]{Table~\ref*{#1}}}
\newcommand\aref[1]{\hyperref[#1]{Appendix~\ref*{#1}}}

\newcommand{\oneline}[1]{%
  \newdimen{\namewidth}%
  \setlength{\namewidth}{\widthof{#1}}%
  \ifthenelse{\lengthtest{\namewidth < \textwidth}}%
  {#1}
  {\resizebox{\textwidth}{!}{#1}}
}
 
\title[Cosmic Rays in AGN Jet-Driven Shocks]{Modeling Cosmic Rays at AGN Jet-Driven Shock Fronts}
 
\author[]{
\parbox[t]{\textwidth}{
Kung-Yi Su$^{1}$\thanks{E-mail: kungyisu@g.harvard.edu}, Greg L. Bryan$^2$,  Philip F. Hopkins$^3$, Priyamvada Natarajan$^{1,4,5}$, Sam B. Ponnada$^3$,  Razieh \ Emami$^6$, Yue Samuel Lu$^7$
}
\vspace*{6pt} \\
$^1$Black Hole Initiative, Harvard University, 20 Garden Street, Cambridge, MA 02138, USA\\
$^2$Department of Astronomy, Columbia University, 550 West 120th Street, New York, NY 10027, USA\\
$^3$TAPIR 350-17, California Institute of Technology, 1200 E. California Boulevard, Pasadena, CA 91125, USA\\
$^4$Department of Astronomy, Yale University, Kline Tower, 266 Whitney Avenue, New Haven, CT 06511, USA\\
$^5$Department of Physics, Yale University, P.O. Box 208121, New Haven, CT 06520, USA\\
$^6$Center for Astrophysics $\vert$ Harvard \& Smithsonian, 60 Garden Street, Cambridge, MA 02138, USA\\
$^7$Department of Astronomy and Astrophysics, University of California San Diego, 9500 Gilman Dr, La Jolla, CA 92093, USA\\
}
\usepackage[all]{hypcap}

\begin{document}
\long\def\/*#1*/{}
\date{Submitted to MNRAS}

\pagerange{\pageref{firstpage}--\pageref{lastpage}} \pubyear{2021}

\maketitle

\label{firstpage}

\begin{abstract}
Feedback from Active Galactic Nuclei (AGN) is a key physical mechanism  proposed to regulate galaxy formation and suppress star formation, primarily in massive galaxies. In particular, cosmic rays (CRs) associated with AGN jets have the potential to efficiently suppress cooling flows and quench star formation. The locus of cosmic ray production and their coupling to gas play a crucial role in the overall self-regulation process. To investigate this in detail, we conduct high-resolution, non-cosmological MHD simulations of a massive $10^{14} {\rm M_\odot}$ halo using the FIRE-2 (Feedback In Realistic Environments) stellar feedback model. We explore a variety of AGN jet feedback scenarios with cosmic rays, examining different values for the cosmic ray energy fraction in jets, cosmic ray coupling sites (in the vicinity of the black hole versus at the shock fronts of large-scale jet cocoons), and jet precession parameters. Our findings indicate that when cosmic rays are injected near the vicinity of the black hole, they efficiently inhibit black hole accretion by suppressing the density before the jet propagates out to large radii. As a result, this leads to episodic black hole accretion, with the jet not having sufficient energy flux to reach large radii and impact cooling flows. Conversely, if the cosmic rays are injected at the shock front of the jet cocoon, not only does the jet sustain a higher overall energy flux for an extended period, but it also disperses cosmic rays out to larger radii, more effectively suppressing the cooling flow. Furthermore, the period and angle of jet precession can influence the position of shock fronts. We identify an optimal range of jet precession periods ($\sim$ tens of Myr) that generates shocks at the inner circumgalactic medium (CGM), where cooling flows are most severe. We report that this specific configuration offers the most effective scenario for cosmic rays at the shock front to suppress the cooling flow and quench star formation.

\end{abstract}


\section{Introduction}
\label{S:intro}
For decades, one of the major unsolved challenges in galaxy formation has been understanding how to effectively ``quench'' massive galaxies (those with stellar masses $\gtrsim 10^{11}{\rm M}{\odot}$, or brighter than $\sim L{\ast}$ in the galaxy luminosity function) and maintain them as ``red and dead" over significant portions of cosmic time as observations have revealed \citep[see, e.g.,][]{2003ApJS..149..289B,2003MNRAS.341...54K,2003MNRAS.343..871M,2004ApJ...600..681B,keres+2005,2005ApJ...629..143B,2006MNRAS.368....2D,2009MNRAS.396.2332K,2010A&A...523A..13P,2012MNRAS.424..232W,2015MNRAS.446.1939F,2015Natur.519..203V}. This issue is intricately tied to the ``cooling flow'' problem. X-ray observations reveal that the hot gas within elliptical galaxies and clusters cools radiatively, with cooling timescales shorter than a Hubble time \citep{1994ApJ...436L..63F,2006PhR...427....1P,2019MNRAS.488.2549S}. However, despite the significant cooling flows observed (reaching up to $\sim 1000 {\rm M}_\odot {\rm yr}^{-1}$ in some clusters), there is a clear discrepancy: neither sufficient amounts of cold gas, as seen from H{\scriptsize I} and CO observations \citep{2011ApJ...731...33M,2013ApJ...767..153W}, nor enough star formation activity \citep{2001A&A...365L..87T,2008ApJ...681.1035O,2008ApJ...687..899R} are detected in these galaxies. Instead, massive galaxies are observed to be red and dead. Simulations and semi-analytic models that do not account for cooling flow suppression, and instead allow gas to cool into the galactic core, generally predict star formation rates (SFRs) over an order of magnitude higher than those observed \citep[for recent examples, see the weak/no-feedback runs in][]{2007MNRAS.380..877S,somerville:2008,2009MNRAS.398...53B,2015MNRAS.449.4105C,2015ApJ...811...73L,2017MNRAS.472L.109A}.

A source of heat or pressure support is essential to counterbalance the observed cooling. Additionally, this heating mechanism must maintain the cool core structure (e.g., the density and entropy profiles) seen in many groups and clusters \citep{1998MNRAS.298..416P,2009A&A...501..835M}. In previous research work, we found that non-AGN feedback mechanisms proposed in the literature, such as stellar feedback from shock-heated AGB winds, Type Ia supernovae (SNe), SNe-injected cosmic rays, magnetic fields, thermal conduction in the circumgalactic medium (CGM) or intra-cluster medium (ICM), and ``morphological quenching" are insufficient to resolve the cooling flow problem \citep[][hereafter \citetalias{2019MNRAS.487.4393S}]{2019MNRAS.487.4393S}. As a result, AGN feedback appears to be the most promising solution to this issue, and extensive theoretical work has been conducted on the subject. For recent investigations into AGN jet feedback, see references noted in later paragraphs, and for other types of AGN feedback, we refer to studies such as \citealt[][]{2017ApJ...837..149G,2017MNRAS.468..751E,2018MNRAS.479.4056W,2018ApJ...866...70L,2018ApJ...856..115P,2018ApJ...864....6Y}, as well as earlier work \citep[e.g.,][]{1998A&A...331L...1S,1999MNRAS.308L..39F,2001ApJ...551..131C,2005ApJ...630..705H,2006ApJS..163....1H,2006MNRAS.365...11C,2009ApJ...699...89C,2012ApJ...754..125C}.

Observational studies also suggest that the energy available from AGN is sufficient to match the cooling rates observed in clusters and groups \citep{2004ApJ...607..800B}. Furthermore, there are clear observations of AGN expelling gas from galaxies and injecting thermal energy via shocks, sound waves, photoionization, Compton heating, or ``stirring" the CGM and ICM. Some of these processes create ``bubbles" of hot plasma, often with significant relativistic components, which are commonly observed around massive galaxies \citep[for a detailed review, see][]{2012ARA&A..50..455F,2018ARA&A..56..625H}.

Despite the plausibility of AGN feedback and the substantial body of work on the subject, the detailed physics behind it remains unclear, as do the precise ``input parameters'' required. Several studies have also pointed out that certain types of AGN feedback models struggle to consistently quench star formation, self-regulate, or meet key observational constraints \citep[e.g.,][]{2004ApJ...607..800B,2006ApJ...645...83V,2020arXiv200400021G,2020MNRAS.491.1190S}. To address these uncertainties, we conducted a comprehensive and systematic investigation of AGN feedback models to determine which, if any, are most viable for solving the cooling flow problem. In previous treatments \cite{2020MNRAS.491.1190S,2021MNRAS.507..175S,2024MNRAS.532.2724S}, we examined various AGN feedback models that assume constant energy injection. Our findings indicated that successful AGN models for halos exceeding $10^{12}{\rm M_\odot}$ feature energy fluxes comparable to the free-fall energy flux at the cooling radius and generate sufficiently large and wide AGN cocoons (AGN heated regions) with long cooling times. Notably, cosmic-ray-dominated models emerged as the most effective at quenching star formation while maintaining stable cool-core, low-SFR halos over extended periods. These models also successfully matched observational constraints on halo gas properties and remained within plausible energy budgets for low-luminosity AGN in massive galaxies.

Although cosmic rays in AGN feedback have been proven capable of quenching massive galaxies, there are still a few gaps in our understanding of how they do so. Cosmic rays are associated with shocks at various scales. The traditional method of injecting cosmic rays around the black hole's vicinity models cosmic rays generated by sub-resolution physics, but does not necessarily track those accelerated by large-scale jet-driven shock fronts. Another limitation of our previous work is that we used a constant energy flux to study the effects of energy injection without modeling black hole accretion, preventing us from capturing the self-regulation of feedback. In this work, we focus on addressing these two important aspects that were not addressed previously. 

 Specifically, we present the first Lagrangian method for injecting cosmic rays at large-scale jet-driven shock fronts using a particle spawning model. We investigate how cosmic ray injection at these shock fronts influences cooling flows and star formation. Additionally, we explore how this process interacts with jet precession, the jet opening angle, the jet velocity, and black hole accretion.

 In \sref{S:methods}, we summarize our initial conditions (ICs), the AGN jet parameters we explore, and the details of our numerical simulations.  In \sref{S:results_const}, we present the results from the constant AGN jet energy flux runs. Following that, we discuss the live accretion runs in \sref{S:results_live}. In \sref{s:discussion}. we address the required cosmic ray energy flux for quenching, the implications for gamma-ray fluxes, and the limitations of this improved model. Finally, we conclude in \sref{s:conclusions}.

\section{Methodology} \label{S:methods}
We perform simulations of isolated galaxies with a dark matter halo mass of approximately $10^{14}{\rm M}_\odot$. The initial conditions are based on observed profiles of cool-core galaxy clusters at low redshift, as described in \sref{S:ic}. Without AGN feedback, despite the galaxies starting with properties consistent with observations, their cooling flow rates and star formation rates (SFRs) rapidly increase, surpassing the observed values for quenched populations by several orders of magnitude \citep{2019MNRAS.487.4393S,2020MNRAS.491.1190S}. Here, we evolve these simulations using various AGN jet models to evaluate their ability to suppress cooling flows and sustain quenched galaxies over time.

Our simulations utilize the {\sc GIZMO} code\footnote{A public version of the code is available at \href{http://www.tapir.caltech.edu/~phopkins/Site/GIZMO.html}{\textit{http://www.tapir.caltech.edu/$\sim$phopkins/Site/GIZMO.html}}} \citep{2015MNRAS.450...53H}, specifically in its meshless finite mass (MFM) mode. This Lagrangian, mesh-free Godunov method combines the strengths of grid-based and smoothed-particle hydrodynamics (SPH) techniques. The code's implementation and extensive tests for hydrodynamics, self-gravity \citep{2015MNRAS.450...53H}, magnetohydrodynamics (MHD) \citep{2016MNRAS.455...51H,2015arXiv150907877H}, anisotropic conduction and viscosity \citep{2017MNRAS.466.3387H,2017MNRAS.471..144S}, and cosmic rays \citep{chan:2018.cosmicray.fire.gammaray} are documented in a series of method papers.

All simulations use the FIRE-2 (Feedback In Realistic Environments) model, incorporating the physical processes of the interstellar medium (ISM), star formation, and stellar feedback. The full details and extensive numerical tests are provided in \citet{2017arXiv170206148H}. Cooling is modeled over the temperature range $10-10^{10}$ K, accounting for various processes including photoelectric and photoionization heating, and cooling via collisional, Compton, fine structure, recombination, atomic, and molecular mechanisms.

Star formation is handled via a sink particle method, restricted to molecular, self-shielded, and locally self-gravitating gas with densities above $n>1000\,{\rm cm^{-3}}$ \citep{2013MNRAS.432.2647H}. Formed star particles represent single stellar populations with metallicities inherited from their progenitor gas particles. Feedback rates for supernovae, mass loss, spectra, and other factors are IMF-averaged values from {\small STARBURST99} \citep{1999ApJS..123....3L}, using a \citet{2002Sci...295...82K} initial mass function (IMF). The stellar feedback model includes: (1) radiative feedback from photoionization and photoelectric heating, as well as radiation pressure across five bands (ionizing, FUV, NUV, optical-NIR, IR), (2) continuous mass loss and energy injection from OB and AGB winds, and (3) SNe Type II and Type Ia, including both prompt and delayed populations, injecting mass, metals, momentum, and energy into the surrounding gas. All simulations, except the `B0' run, also include MHD, and fully anisotropic Spitzer-Braginski conduction and viscosity.

To briefly summarize the included cosmic ray physics, we track a single energy bin of $\sim$GeV protons, which dominate the cosmic ray pressure, in the ultra-relativistic limit. Cosmic ray transport includes streaming at the local Alfvén speed, with an appropriate streaming loss term that thermalizes according to \citet{Uhlig2012}, where $v_{\rm st}=v_A$. Additionally, diffusion is modeled with a fixed diffusivity, $\kappa_{\rm CR}$, and the system accounts for adiabatic energy exchange between gas and cosmic ray pressure, as well as hadronic and Coulomb losses \citep{2008MNRAS.384..251G}. Streaming and diffusion occur anisotropically along magnetic field lines. Previous studies \citep{chan:2018.cosmicray.fire.gammaray, hopkins:cr.mhd.fire2, 2021MNRAS.501.4184H} demonstrated that matching observed $\gamma$-ray luminosities requires a diffusivity of $\kappa_{\rm CR}\sim 10^{29}\,{\rm cm^{2}\,s^{-1}}$, which is consistent with detailed cosmic ray transport models incorporating an extended gaseous halo around the galaxy \citep[e.g.,][]{1998ApJ...509..212S, 2010ApJ...722L..58S, 2011ApJ...729..106T}. Thus, we adopt this as our fiducial value \footnote{Theoretically motivated cosmic ray transport models, such as extrinsic turbulence and self-confinement, have been explored in previous studies \cite[e.g.,][]{2022MNRAS.517.5413H,2024MNRAS.530L...1P}. However, \cite{2022MNRAS.517.5413H} demonstrates that these models struggle to accurately reproduce the observed cosmic ray spectrum. In this work, we adopt the simplest assumption of constant diffusivity.}. For the simulation that includes cosmic rays from supernovae, 10\% of the supernova energy is allocated to cosmic rays.

\subsection{Initial conditions}
\label{S:ic}

The initial conditions applied in this work are comprehensively outlined in \citet{2019MNRAS.487.4393S,2020MNRAS.491.1190S,2021MNRAS.507..175S,2024MNRAS.532.2724S}. Our galaxy model extends out to a distance of at least three times the virial radius. The initial setup is specifically designed to closely replicate observed cool-core systems with a halo mass of $\sim 10^{14}M_\odot$ at $z \sim 0$ \citep[see, e.g.,][]{2012ApJ...748...11H,2013MNRAS.436.2879H,2013ApJ...775...89S,2015ApJ...805..104S,2017A&A...603A..80M}, and the relevant parameters are summarized in \tref{tab:ic}. The dark matter (DM) halo, bulge, black hole, and the gas and stellar disk are initialized with live particles, following the procedures described by \cite{1999MNRAS.307..162S} and \cite{2000MNRAS.312..859S}.

We employ a spherical Navarro-Frenk-White (NFW) profile \citep{1996ApJ...462..563N} for the DM halo, a \cite{1990ApJ...356..359H} model for the stellar bulge, and an exponentially declining, rotationally supported disk of gas and stars, initialized with a Toomre parameter of $Q\approx1$ for the baryonic component. The black hole (BH) is assigned a mass equivalent to approximately $1/300$ of the bulge mass \citep[e.g.,][]{2004ApJ...604L..89H}. The gas halo, modeled with a $\beta$-profile, rotates at twice the angular momentum of the dark matter halo. About 10-15\% of the gravitational support is provided by this rotation, while the bulk of the support comes from thermal pressure, consistent with what is expected in large halos. Once again, the values of the adopted parameters are summarized in \tref{tab:ic}.

The initial metallicity of the CGM/ICM decreases with radius, from solar metallicity ($Z=0.02$) to $Z=0.001$, following the relation $Z(r)=0.02\,(0.05+0.95/(1+(r/r_Z)^{1.5}))$, where $r_Z=20$ kpc. The initial magnetic field is azimuthal, with a seed value taken to be $|{\bf B}|=B_0/(1+(r/r_B)^{2})^{\beta_B}$, which is later amplified. This field extends throughout the ICM, with $B_0=0.3 \, {\rm \mu G}$, $r_B=20$ kpc, and $\beta_B=0.375$. The initial cosmic ray energy density is set in equi-partition with the local magnetic energy density, though both remain weak compared to the thermal energy. These initial conditions are run adiabatically (without cooling or star formation) for 50 Myr to allow for the dissipation of any initial transients.

Our initial conditions produce X-ray luminosities in the 0.5–7 keV band of $\sim 3 \times 10^{43}~{\rm erg,s}^{-1}$ \citep{2024MNRAS.532.2724S}, which are consistent with observational data \citep{2002ApJ...567..716R,2006ApJ...648..956S}. We note that a resolution study is presented in the appendix of \citet{2019MNRAS.487.4393S}. To improve convergence, we employ a hierarchical super-Lagrangian refinement technique \citep{2019MNRAS.487.4393S,2020MNRAS.491.1190S} to achieve a mass resolution of approximately $3 \times 10^4 \, {\rm M}{\sun}$ in the core region and along the z-axis where the jet is introduced. This resolution is significantly higher than that of most previous global simulations. The mass resolution decreases with both radius ($r_{\rm 3d}$) and distance from the z-axis ($r_{\rm 2d}$), scaling roughly as $r_{\rm 3d}$ and $2^{r_{\rm 2d}/10 \, {\rm kpc}}$, down to a minimum resolution (maximum mass) of $2 \times 10^6 \, {\rm M}{\sun}$. The finest resolution is focused in regions where either $r_{\rm 3d}$ or $r_{\rm 2d}$ is less than 10 kpc.

\begin{table*}
\begin{center}
 \caption{Properties of Initial Conditions for the Simulations/Halos Studied In This Paper}
 \label{tab:ic}
 \resizebox{17.7cm}{!}{
\begin{tabular}{l|c|cc|ccc|c|cc|cc|cc|ccc}
 \hline
 \hline
&&\multicolumn{2}{c|}{{Resolution}}&\multicolumn{3}{c|}{{DM halo}}&&\multicolumn{2}{c|}{{Stellar Bulge}}&\multicolumn{2}{c|}{{Stellar Disc}}&\multicolumn{2}{c|}{{Gas Disc}}&\multicolumn{3}{c}{{Gas Halo}}\\
\hline
$\,\,\,\,$Model & $R_{\rm 200}$ &$\epsilon_g^{\rm min}$ &$m_g$        &$M_{\rm DM}$   &$r_{\rm dh}$            &$\rho_0$    &$M_{\rm baryon}$    &$M_b$ 
                 &$a$          &$M_d$        & $r_d$             &$M_{\rm gd}$       &$r_{\rm gd}$         &$M_{\rm gh}$         &$r_{\rm gh}$ &$\beta$    \\
                 &(kpc)  &(pc)        &(M$_\odot$)  &(M$_\odot$)      & (kpc)           &(g/cm$^3$)           &(M$_\odot$)      &(M$_\odot$) 
                  &(kpc)        &(M$_\odot$)  &(kpc)            &(M$_\odot$)    &(kpc)            &(M$_\odot$)      &(kpc)\\
\hline
&&&&&&&&&&&&&\\
{ $\,\,\,\,$m14} &{880} &{1}  &{3e4}      &{ 6.7e13}      &{ 220}       &{ 4.3e-26}         &{ 9.6e12}        &{2.0e11}
                     &{3.9}     &{2.0e10}          &{3.9}              &{1.0e10}           &{3.9}              &{9.4e12}           &{ 22}    &{ 0.5}          \\          
   
\hline 
\hline
\end{tabular}
}
\end{center}
\begin{flushleft}
Parameters of the galaxy/halo model studied in this work (\sref{S:ic}): 
(1) Model name. The number following `m' labels the approximate logarithmic halo mass. 
(2) $R_{\rm 200}$. The radius that encloses an average density of 200 times the critical density.
(3) $\epsilon_g^{\rm min}$: Minimum gravitational force softening for gas (the softening for gas in all simulations is adaptive, and matched to the hydrodynamic resolution; here, we quote the minimum Plummer equivalent softening).
(4) $m_g$: Gas mass (resolution element). There is a resolution gradient for m14, so its $m_g$ is the mass of the highest resolution elements.
(5) $M_{\rm DM}$: Dark matter halo mass. 
(6) $r_{\rm dh}$: NFW halo scale radius.
(7) $\rho_0$: Density normalization of NFW halo.
(8) $M_{\rm baryon}$: Total baryonic mass. 
(9) $M_b$: Bulge mass.
(10) $a$: Bulge Hernquist-profile scale-length.
(11) $M_d$ : Stellar disc mass.
(12) $r_d$ : Stellar disc exponential scale-length.
(13) $M_{\rm gd}$: Gas disc mass. 
(14) $r_{\rm gd}$: Gas disc exponential scale-length.
(15) $M_{\rm gh}$: Hydrostatic gas halo mass. 
(16) $r_{\rm gh}$: Hydrostatic gas halo $\beta$ profile scale-length.
(17) $\beta$: Hydrostatic gas halo $\beta$ profile value.
\end{flushleft}
\end{table*}

\subsection{Black hole accretion}\label{S:accretion}

For the  BH accretion we utilize a gravitational torque derived growth mode following \cite{2010MNRAS.407.1529H} and \cite{2011MNRAS.415.1027H}. The model assumes accretion is regulated by gravitational torques from asymmetries in the gravitational potential including collisionless interactions with (dark matter and star) particles and the gas, and gaseous self-interaction. The accretion rate is given by:
\begin{align}\label{eq:gravt}
\dot{M}_{\rm acc}=C_{\rm acc} \frac{(M_{\rm BH}/M_d)^{1/6}M_{\rm gas}\Omega}{1+3(M_{d}/10^9 \,{\rm M}_\odot)^{1/3}(M_{\rm gas}/M_d)},
\end{align}
where  $M_{\rm gas}$ and $M_d$ are the total gas mass and the total mass of angular-momentum-supported material in the black hole kernel; $\Omega=\sqrt{G M_{\rm tot}/R^3}$ is the dynamical frequency at the force-softening radius, enclosing a total mass of $M_{\rm tot}$.
We also account for unresolved stellar feedback effects on the accretion kernel scale by assuming stellar feedback decouples a fraction of the material, following \cite{2019MNRAS.483.5548G,2022MNRAS.510..630H}. As a result, the accretion efficiency is modified as:
\begin{align}
C_{\rm acc} = \frac{a_{\rm g,eff}}{a_{\rm g,crit}+a_{\rm g,eff}},
\end{align}
where $a_{\rm g,eff}$ is the gravitational acceleration at the black hole kernel scale and $a_{\rm g,crit} \sim 10^{-7} \,{\rm cm \, s^{-2}}$.

The accreted mass is first added to a sub-grid alpha disk, which acts as a reservoir of mass $M_\alpha$. The black hole is then fed with $\dot{M}_{\rm BH} = M_\alpha/t_\alpha$, where the accretion timescale is given by:
\begin{align}
t_\alpha =42 {\rm Myr} \left(\frac{1+M_{\rm BH}}{M_\alpha}\right)^{0.4}.
\end{align}

\subsection{The AGN jet and cosmic rays}

We launch the jet with a particle spawning method, which creates new gas cells (resolution elements) from the central black hole. With this method, we have better control of the jet properties as the launching is less dependent on the neighbor-finding results. We can also enforce a higher resolution for the jet elements. The numerical method adopted in this work is similar to \cite{2020MNRAS.497.5292T}, which studied the effects of broad absorption line (BAL) wind feedback on disk galaxies. The spawned gas particles have a mass resolution of 5000 ${\rm M}_\odot$ and are forbidden to de-refine (merge into a regular gas element) before they decelerate to 10\% of the launch velocity. Two particles are spawned in opposite jet directions at the same time when the accumulated jet mass flux reaches twice the target spawned particle mass, so linear momentum is always exactly conserved. Initially, the spawned particle is randomly placed on a sphere with a radius of $r_0$, which is either 10 pc or half the distance between the black hole and the closest gas particle, whichever is smaller. If the particle is initialized at a position $(r_0,\theta_0,\phi_0)$ and the jet opening angle of a specific model is $\theta_{\rm op}$, the initial velocity direction of the jet will be set at $2\theta_{\rm op}\theta_0/\pi$ for $\theta_0<\pi/2$ and at $\pi-2\theta_{\rm op}(\pi-\theta_0)/\pi$ for $\theta_0>\pi/2$. With this, the projected paths of any two particles will not intersect. We have $\theta_0=1^o$ for all runs in this work. For the runs without jet precession, the jet direction is at the z axis. For the run with jet precession, the jet precesses around the z axis with a precession angle of $\theta_{\rm p}=45^o$, and with precession period $t_{p}$.

For the spawned jet particles, we initialize their velocity ($V_{\rm jet}$), temperature ($T_{\rm jet}$), and magnetic field ($B_{\rm jet}$). For the cosmic rays, we consider two cases: cosmic rays that are injected near the black hole and cosmic rays injected at jet-driven shock fronts on large scales.

\begin{itemize} 
\item {\bf Cosmic rays near the black hole vicinity:$\,\,\,$} The motivation for injecting cosmic rays in the vicinity of the black hole is to model cosmic rays accelerated by sub-resolution (sbugrid) physics. To achieve this, we assign a fixed cosmic ray energy per unit mass $(e_{\rm CR,\,BH})$ to the newly spawned jet particles.

\item {\bf Cosmic rays at the large-scale jet-driven shock front:$\,\,\,$} The motivation here is to model cosmic rays accelerated at the primary jet-driven shock via Fermi acceleration. As a simple model for this process, we inject cosmic rays when the spawned jet particles decelerate to $1/4$ of their initial velocity, at which point they have almost surely encountered a strong shock \footnote{Assuming an adiabatic index of $\gamma = 5/3$ and strong shock conditions, the post-shock velocity is reduced to 1/4 of the pre-shock velocity.}. The cosmic ray energy injected at that point is, once again, implemented with a fixed energy per unit mass $(e_{\rm CR,\,jet})$. 
\end{itemize}

For the runs with constant jet fluxes, we also keep a constant mass flux of jet ($\dot{M}_{\rm jet}$). For the runs with live accretion, as described in \sref{S:accretion}, the mass flux of the jet is tied to the black hole accretion rate as follows:
\begin{align}
\dot{M}_{\rm jet}=\eta_{\rm jet,\, mass} \dot{M}_{\rm BH},
\end{align}
where $\eta_{\rm jet,\, mass}$ is the constant feedback mass loading.

With the prescription above, the net energy flux from the jet is:
\begin{align}
\dot{E}_{\rm jet}&=\dot{M}_{\rm jet} \left( \frac{1}{2} V_{\rm jet}^2+ \frac{3 k T_{\rm jet}}{2 \mu m_p}+\frac{B_{\rm jet}^2}{2\rho} +e_{\rm CR,\,BH}+e_{\rm CR,\,jet}\right),\notag\\
&\equiv\dot{M}_{\rm BH} c^2 \left( \eta_{\rm kin} + \eta_{\rm th} + \eta_{\rm mag} + \eta_{\rm CR,\,BH} + \eta_{\rm CR,\,jet} \right)
\end{align}
where $\mu$ is the averaged particle mass with respect to the proton mass, $m_p$ is the proton mass and $\rho$ is the density at the black hole vicinity.

We conduct a set of runs with constant energy flux to explore the impact of injecting cosmic rays at the large-scale jet-driven shock front and how this interacts with other jet parameters. The list of runs and their parameters is provided in \tref{tab:run_const}.
In addition, we performed a more extensive set of runs to test how the previously described variations couple with black hole accretion. The list of these runs and their parameters is provided in \tref{tab:run_live}.

\setlength{\tabcolsep}{4pt}
\begin{table*}
\begin{center}
 \caption{Physics variations for constant-flux runs}
 \label{tab:run_const}
\resizebox{17.7cm}{!}{%
\begin{tabular}{lc|cc|ccccc|ccc|cc|ccc}
\hline
\hline
\multicolumn{2}{c|}{}&\multicolumn{2}{c|}{Results}&\multicolumn{8}{c|}{Input Jet Fluxes}&\multicolumn{4}{c}{Other Jet Parameters} \\
\hline
Model           &$\Delta t$& SFR &Summary  &$\dot{E}_{\rm kin}$ &$\dot{E}_{\rm th}$ &$\dot{E}_{\rm mag}$  &$\dot{E}_{\rm CR, BH}$&$\dot{E}_{\rm CR, jet}$ &$\dot{M}$ &v &$\dot{P}$ &T & B &$\theta_{\rm op}$ &$\theta_{\rm p}$ & $T_{p}$ \\
                &Gyr  &\multicolumn{2}{l|}{${\rm M}_\odot$ yr$^{-1}$}&\multicolumn{5}{c|}{erg s$^{-1}$} & ${\rm M}_\odot$ yr$^{-1}$ &km s$^{-1}$ &cgs &K &G & deg&deg & {\rm Myr}\\
\hline

NoJet              &1.5 & 42 &  CF &\multicolumn{5}{c|}{N/A} &\multicolumn{3}{c|}{N/A} &\multicolumn{2}{c|}{N/A} &\multicolumn{3}{c}{N/A} \\ 

fixed-v1e4-CRBH0.1-pr100&1  &3.5&quenched&5.8e43& 1.9e41&2.3e41&5.8e42&0&2&9.5e3&3.9e34&1e7&1e-3&1&45&100\\ 
fixed-v1e4-CRshock0.1&0.74       &26&CF&5.8e43& 1.9e41&1.2e41&0&5.8e42&2&9.5e3&3.9e34&1e7&1e-3&1&N/A&N/A\\ 
fixed-v1e4-CRshock0.1-iso&1   &1.6&quenched&5.8e43& 1.9e41&2.1e41&0&5.8e42&2&9.5e3&3.9e34&1e7&1e-3&90&45&100\\ 
fixed-v1e4-CRshock0.1-pr100&1 &1.3&quenched&5.8e43& 1.9e41&2.3e41&0&5.8e42&2&9.5e3&3.9e34&1e7&1e-3&1&45&100\\ 
fixed-v1e4-CRshock0.1-pr10&1  &2.7&quenched&5.8e43& 1.9e41&2.3e41&0&5.8e42&2&9.5e3&3.9e34&1e7&1e-3&1&45&10\\ 

\hline 
\hline
\end{tabular}
}
\end{center}
\begin{flushleft}
Columns list: 
(1) Model name: The naming of each model starts with fixed or live, indicating whether it has live accretion. It is then followed by the type of CR jet, with CR launched at the BH vicinity `CRBH' or at the jet-driven shock front `CRshock'. Next is the fraction of energy in cosmic rays. All runs are with small open-angle jets, except for the run labeled `iso', which has an isotropic wind. The number following pr indicates the processing period in Myr.
(2) $\Delta t$: Simulation duration. All simulations are run to $1\,$Gyr, unless the halo is completely completely unaffected.
(3) The SFR averaged over the 0.8-1 Gyr (or the last 0.2 Gyr).
(4) Summary of the results:  `CF' and `quenched' correspond, respectively, to an SFR of $\gtrsim 5 {\rm M_\odot}$ yr$^{-1}$ and $\lesssim 5 {\rm M_\odot}$ yr$^{-1}$. 
(5, 6, 7) $\dot{E}_{\rm kin}$, $\dot{E}_{\rm th}$, and $\dot{E}_{\rm mag}$ tabulate the total energy input of the corresponding form. The magnetic energy flux is the time-averaged value.
(8) $\dot{E}_{\rm CR,\,BH}$ and $\dot{E}_{\rm CR,\,jet}$: The CR energy flux launched around a black hole or at the large-scale jet-driven shock front.
(9, 10, 11) $\dot{M}$, $v$, and $\dot{P}$ tabulate the mass flux, jet velocity, and momentum flux, respectively.
(12) T: The initial temperature of the jet.
(13) B: The maximum initial magnetic field strength of the jet.
(14) $\theta{op}$: The opening angle of the jet or wind.
(15) $\theta_{p}$: The precession angle of the jet or wind.
(16) $T_{p}$: Precession period.
\end{flushleft}
\end{table*}
\setlength{\tabcolsep}{6pt}

\setlength{\tabcolsep}{4pt}
\begin{table*}
\begin{center}
 \caption{Physics variations for live accretion runs}
 \label{tab:run_live}
\resizebox{17.7cm}{!}{%
\begin{tabular}{lc|c|ccccc|c|ccc|ccc|c}
\hline
\hline
\multicolumn{2}{c|}{}&\multicolumn{1}{c|}{Results}&\multicolumn{6}{c|}{Input Jet Fluxes}&\multicolumn{6}{c|}{Other Jet Parameters} & CR in SNe \\
\hline
Model           &$\Delta t$ &Summary  &$\eta_{\rm kin}$ &$\eta_{\rm th}$ &$\eta_{\rm mag}$  &$\eta_{\rm CR, BH}$&$\eta_{\rm CR, jet}$ &$\eta_{\rm jet,\, mass}$ &v  &T & B &$\theta_{\rm op}$ &$\theta_{\rm p}$ & $T_{p}$ & $f_{\rm CR,\,SNe}$\\
                &Gyr  &\multicolumn{1}{l|}{}&\multicolumn{5}{c|}{}& &km s$^{-1}$ &K &G & deg&deg & {\rm Myr}\\
\hline

NoJet              &1.5  &  CF &\multicolumn{5}{c|}{N/A}&N/A &\multicolumn{3}{c|}{N/A} &\multicolumn{3}{c|}{N/A} &0 \\ 
live-v1e4-CRBH0.1-pr100       &1.2&slight $\downarrow$&5e-3&1.6e-5&4.0e-6 (2.0e-7)&5e-4&0&9&1e4&1e7&1e-4&1&45&100&0\\
live-v1e4-CRshock0.1            &1.2&CF&5e-3&1.6e-5&1.7e-6 (2.2e-7)&0&5e-4&9&1e4&1e7&1e-4&1&N/A&N/A&0\\
live-v1e4-CRshock0.1-pr100      &2& slight $\downarrow$&5e-3&1.6e-5&4.9e-5 (8.8e-7)&0&5e-4&9&1e4&1e7&1e-4&1&45&100&0\\
live-v1e4-CRshock0.1-pr10       &1.1&slight $\downarrow$&5e-3&1.6e-5&1.6e-5 (1.5e-6)&0&5e-4&9&1e4&1e7&1e-4&1&45&10&0\\
live-v1e4-CRshock0.3-pr100      &2.8&episodic q&5e-3&1.6e-5&2.4e-4 (3.1e-6)&0&1.5e-3&9&1e4&1e7&1e-4&1&45&100&0\\
live-v1e4-CRshock0.3-pr10       &1&episodic  q&5e-3&1.6e-5&4.3e-5 (6.0e-6)&0&1.5e-3&9&1e4&1e7&1e-4&1&45&10&0\\ 
live-v3e4-CRshock0.1-pr10       &0.85&CF&4.5e-2&1.6e-5&3.6e-4 (9.4e-4)&0&4.5e-3&9&3e4&1e7&1e-4&1&45&10&0\\
live-v1e4-CRshock0.1-CRSNe-pr100&1.4&episodic q&5e-3&1.6e-5&3.4e-5 (7.0e-7)&0&5e-4&9&1e4&1e7&1e-4&1&45&100&0.1\\
live-v1e4-NoCR                     &1.2&slight $\downarrow$&5e-3&1.6e-5&6.3e-8 (8.8e-8)&0&0&9&1e4&1e7&1e-4&1&N/A&N/A&0\\
live-v1e4-NoCR-pr100               &2&slight $\downarrow$&5e-3&1.6e-5&9.9e-6 (8.1e-8)&0&0&9&1e4&1e7&1e-4&1&45&100&0\\ 
live-v3e3-CRBH1              &1&slight $\downarrow$&4.5e-4&1.6e-5&8.9e-5 (4.2e-5)&0&5e-3&9&3e3&1e7&1e-4&1&N/A&N/A&0\\
\hline 
\hline
\end{tabular}
}
\end{center}
\begin{flushleft}
This is a partial list of simulations studied here: each halo was run with jet models and fluxes scaled from the more successful m14 run we concluded in \citetalias{2021MNRAS.507..175S}. Columns list: 
(1) Model name: The naming of each model starts with fixed or live, indicating whether it has live accretion. It is then followed by the type of CR jet, with CR launched at the BH vicinity `CRBH' or at the jet-driven shock front `CRshock' or no CR `NoCR'. Next is the fraction of energy in cosmic rays.`CRSNe' represents the inclusion of cosmic rays from supernovae. All runs are with small open-angle jets. The number following pr indicates the processing period in Myr.
(2) $\Delta t$: Simulation duration. All simulations are run to $1\,$Gyr, unless the halo is completely unaffected.
(3) Summary of the results: `CF' indicates that the cooling flow is barely suppressed. `Episodic q' means that star formation is suppressed to $\lesssim 5 {\rm M_\odot}$ yr$^{-1}$ episodically. `Slight $\downarrow$' includes everything that falls in between.
(4, 5, 6) $\eta_{\rm kin}$, $\eta_{\rm th}$, and $\eta_{\rm mag}$ tabulate the energy efficiency of the corresponding form. The magnetic energy efficiency is expressed as the time-averaged value, while the value in parentheses represents the mass-weighted average.
(7,8) $\eta_{\rm CR,\,BH}$ and $\eta_{\rm CR,\,jet}$: The CR energy efficiency for the CR launched around a black hole or at the large-scale jet-driven shock front.
(9) $\eta_{\rm jet,\,mass}$: The feedback mass loading.
(10)  $v$: Jet velocity.
(11) T: The initial temperature of the jet.
(12) B: The maximum initial magnetic field strength of the jet; (t) and (p) denote toroidal and poloidal, respectively.
(13) $\theta{op}$: The opening angle of the jet or wind.
(14) $\theta_{p}$: The precession angle of the jet or wind.
(15) $T_{p}$: Precession period.
(16) $f_{\rm CR,\,SNe}$: CR energy fraction in SNe.
\end{flushleft}
\end{table*}
\setlength{\tabcolsep}{6pt}

\section{Results for constant-flux-jet runs} \label{S:results_const}

In this section, we discuss the 5 simulations with constant energy flux jets, with varied opening angles, precession periods, and cosmic ray injection sites.

\subsection{The star formation rate for runs with constant flux jets}\label{S:sfr_const}
\begin{figure}
\includegraphics[width=8cm]{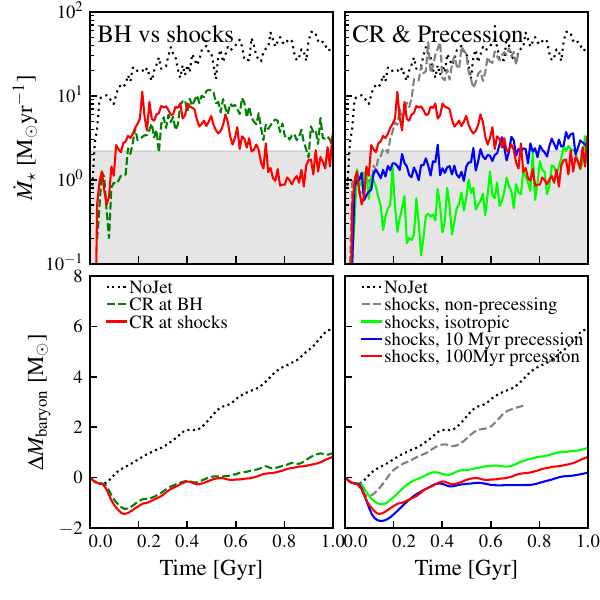}
\label{fig:sft_fixed}
\caption{Star formation rate (top) and core baryonic mass within 30 kpc (bottom) for the constant jet energy flux runs. With the same energy flux, injecting cosmic rays at the jet-driven shock front (CRshock0.1-pr100Myr) results in delayed suppression of star formation compared to injecting cosmic rays near the black hole (CRBH0.1-pr100Myr). Additionally, faster variations in wind direction (`iso', pr10Myr', `pr100Myr' from fastest to slowest) lead to greater suppression of early star formation.}
\end{figure}

The star formation rate and the net change in core baryonic mass for the constant energy flux jet runs are shown in  \fref{fig:sft_fixed}. The baryonic mass accounts for the sum of gas mass and stellar mass within 30 kpc. With a sufficiently high energy flux ($6\times 10^{43}$ erg s$^{-1}$), comparable to the free-fall energy at the cooling radius, both jet models—whether cosmic rays are injected near the black hole or at the large-scale jet-driven shock front—quench star formation and suppress cooling flows. When CRs are injected near the black hole, star formation is suppressed more rapidly. However, when CRs are injected at the large-scale jet-driven shock front, although they start out being slightly less effective at the beginning of the run, the quenching effect picks up and by $\sim$0.5 Gyr, it becomes comparable to the dynamical time at a few tens of kpc, where most of the CRs are deposited. After that point, the jet model with CR injection at the jet-driven shock front becomes slightly more effective at quenching star formation. A similar trend is observed in the panel showing the core baryonic mass. Initially, the jet with CRs injected near the black hole suppresses the core baryonic mass more efficiently. After $\sim0.5$ Gyr, the jet with CRs injected at the large-scale jet-driven shock front becomes slightly more effective in suppressing the cooling flow, resulting in a marginally weaker growth of the core baryonic mass.

In the runs with cosmic rays injected at the large-scale shock front, we also varied the jet precession and jet opening angle. In most of these runs, the energy flux effectively quenches star formation. The only exception is the non-precessing run, where CRs couple at the shock front. In this case, the jet cocoon remains narrow, and the shock front is pushed progressively outward. As a result, a large fraction of the CR energy is deposited at radii over 100 kpc, concentrated within a narrow opening angle, which has little impact on the bulk cooling flows. The CR map can be seen in \fref{fig:morph_CR_const} and will be discussed in \sref{s:cr_profile_const}.

Introducing some degree of precession or increasing the jet's opening angle makes the jet much more efficient at quenching star formation, consistent with our previous findings reported in \cite{2021MNRAS.507..175S}. However, we did not include CRs in that subset of simulations. The run with isotropic winds generates shocks on the smallest scales. The wind itself, along with the injected CRs, significantly affects the ISM gas, leading to the fastest suppression of the star formation rate.

In the two runs with precessing jets, the shorter the precession period is, the more suppressed the early ($<$0.5Gyr) star formation is, and the faster quenching occurs. This is because a longer precession period allows the jet to remain aligned in the same direction for a longer time, pushing the shock front farther out. As the shock front and the region where CRs are deposited move outward, quenching occurs over a longer timescale but at a slower rate. The difference in CR profiles is seen in \fref{fig:pressure_prof_constant} and will be discussed in \sref{s:cr_profile_const}.

It is worth noting that both variations with precessing jets do not suppress star formation as quickly as the isotropic wind case. The isotropic wind run can essentially be seen as a jet precessing randomly with a period of 5,000 years, which corresponds to the time between two particle spawning events given our mass flux (2 ${\rm M_\odot}$ yr$^{-1}$) and the spawned particle mass ($5000 {\rm M_\odot}$). This supports the idea that longer duty cycles result in quenching over larger radii but with a temporally slower effect. After approximately 0.6 Gyr—a sufficiently long time for the jet with the longest precession period (100 Myr) to quench—the star formation rate and the variations in core baryonic mass converge despite earlier differences.

\subsection{Cosmic ray distribution and the pressure gradient for runs with constant flux jets} \label{s:cr_profile_const}

As noted in our previous works \citep{2021MNRAS.507..175S, 2024MNRAS.532.2724S}, the cosmic ray pressure gradient is a key factor in causing cosmic ray jets to quench more efficiently. Here, we examine the distribution of cosmic ray pressure along with the thermal pressure gradient for the constant energy flux runs.

\begin{figure}
\includegraphics[width=8cm]{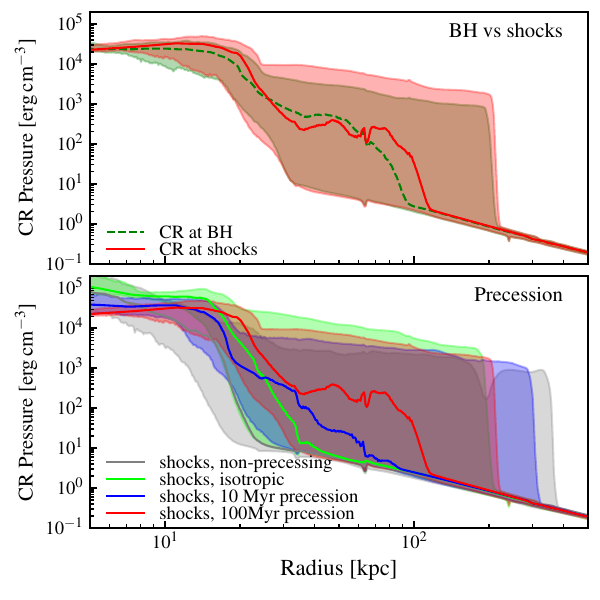}
\label{fig:pressure_prof_constant}
\caption{Mass-weighted median cosmic ray pressure profile (calculated over 0.95–1 Gyr or the last 50 Myr of the run) of the constant jet energy flux runs. The shaded region represents the 2\% and 98\% percentiles at each radius. Injecting cosmic rays at the large-scale jet-driven shock front (CRshock0.1-pr100Myr) produces a slightly more extended cosmic ray distribution compared to injecting cosmic rays near the black hole (CRBH0.1-pr100Myr). Faster changes in feedback direction (from fast to slow: `iso', `pr10Myr', `pr100Myr') lead to more concentrated cosmic ray profiles. In the non-precessing jet case (CRshock0.1), the most extended cosmic rays are visible only in the 98th percentile, as the cosmic rays are concentrated in a small volume with little mass.}
\end{figure}

\begin{figure*}
\includegraphics[width=16cm]{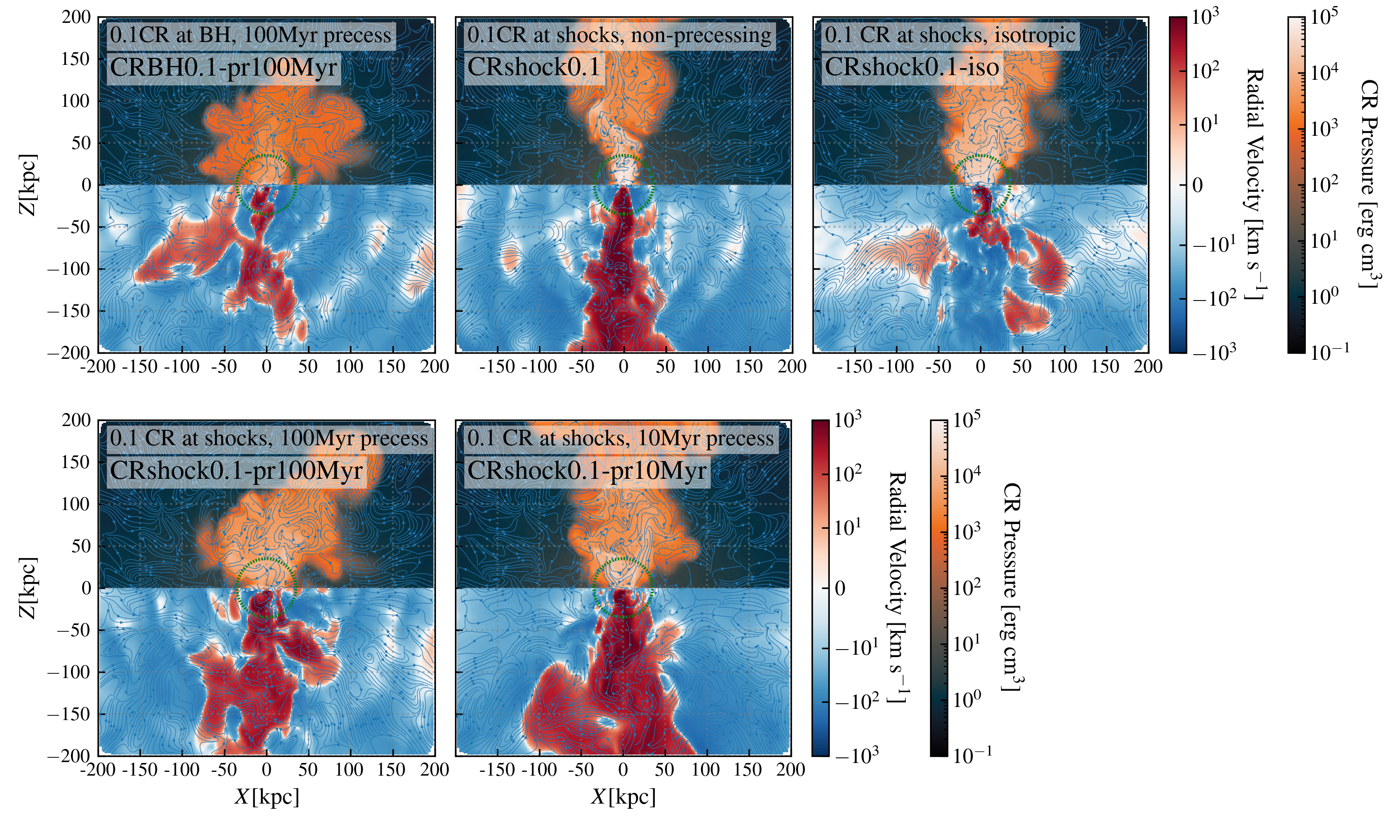}
\label{fig:morph_CR_const}
\caption{Morphological plot of the constant jet energy flux runs for a central slice within $|y|<10$ kpc (at 1 Gyr or at the end of the run).
 The upper half shows the cosmic ray distribution, while the lower half displays the radial velocity. Blue curves represent the magnetic field lines. Injecting cosmic rays at the large-scale jet-driven shock front (CRshock0.1-pr100Myr) results in a slightly more extended cosmic ray distribution compared to injection near the black hole (CRBH0.1-pr100Myr). Faster changes in feedback direction (from fast to slow: `iso', `pr10Myr', `pr100Myr') lead to more concentrated cosmic ray profiles. The non-precessing jet (CRshock0.1) produces the narrowest beam of cosmic rays, reaching the largest radius.}
\end{figure*}

\fref{fig:pressure_prof_constant} shows the radial profiles of mass-weighted median pressure. The shaded region represents the 2-sigma range above and below the median (i.e., the 2\% and 98\% percentiles). In the upper panel, where we compare runs with cosmic rays injected near the black hole and at the large-scale jet-driven shock front, we observe that the latter case results in a more extended cosmic ray distribution. This is evident in the mass-weighted median and even more pronounced in the 98th percentile of the cosmic ray distribution. The cosmic ray morphological distribution is shown more intuitively in \fref{fig:morph_CR_const}.  In the run where cosmic rays are injected at the large-scale shock front (CRshock0.1-pr100Myr', lower left), the cosmic ray distribution is more extended compared to the run with cosmic rays injected near the black hole (CRBH0.1-pr100Myr', upper left).

In the runs where cosmic rays are injected at the large-scale jet-driven shock front without jet precession (CRshock0.1), the mass-weighted median cosmic ray pressure appears lowest at large radii and most concentrated within 10 kpc. This occurs because cosmic rays in this case are confined to a finite volume with a small mass, covering a narrow opening angle, which does not significantly affect the median. However, the 98th percentile (upper edge of the shaded region) reveals that cosmic rays are, in fact, distributed over the largest radii in this run. This is also visually apparent in \fref{fig:morph_CR_const}, where the `CRshock0.1' run exhibits the narrowest jet cocoon with cosmic rays extending to very large radii.

Finally, we compare the runs with isotropic winds and precessing jets with different periods, for when cosmic rays are injected at the shock front. Consistent with the discussion in \sref{S:sfr_const}, the faster the variation in jet direction, from fast to slow, as seen in the runs `CRshock0.1-iso', `CRshock0.1-pr10Myr', and `CRshock0.1-pr100Myr'—the less extended the cosmic ray profile becomes, as reflected in the mass-weighted median.

\begin{figure*}
\includegraphics[width=16cm]{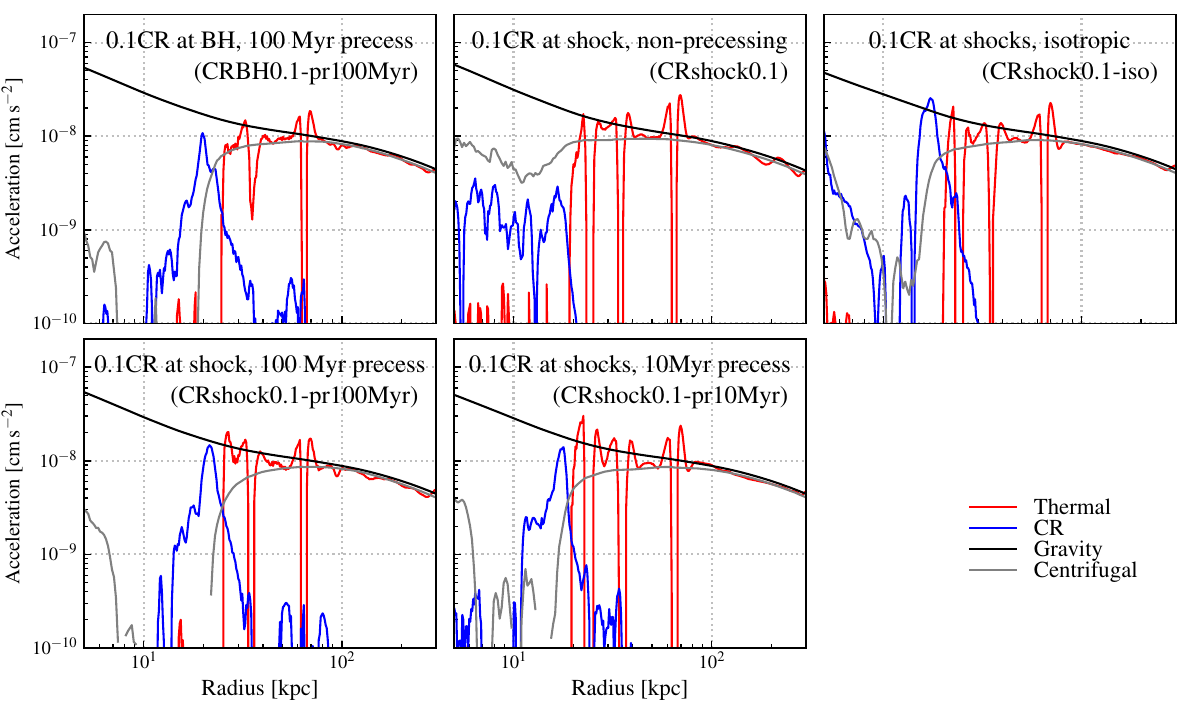}
\label{fig:pressure_gradient_const}
\caption{Comparison of gravitational, rotational, thermal pressure, and CR pressure gradient acceleration in the constant jet energy flux runs (averaged over 0.95–1 Gyr or the last 50 Myr of the run). The centrifugal acceleration is defined as $GM_{\rm enc}/r^2 - v_{\rm rot}^2/r$. In the core region, where cooling is rapid, the thermal pressure gradient does not provide outward support. In all quenched runs (except for the non-precessing jet `CRshock0.1' run), the cosmic ray pressure gradient  balances gravity in the core region.}
\end{figure*}

These variations in cosmic ray profiles significantly affect the cosmic ray pressure distributions, which in turn has a major impact on the suppression of cooling flows.  \fref{fig:pressure_gradient_const} illustrates the pressure-gradient driven acceleration in comparison with the gravitational acceleration and the net inward centrifugal acceleration (gravitational acceleration minus rotational support) for the snapshot after 1 Gyr. At this time, except for the non-precessing run, the pressure gradient in all other runs is comparable to the centrifugal acceleration at 10-30 kpc, indicating a significant effect. Notably, the radius where the cosmic ray pressure gradient is most important is around 10-30 kpc, which is roughly around the cooling radius of this system. At this radius, radiative cooling is efficient and the thermal pressure gradient (red line) is not effective. Beyond this radius, the thermal pressure gradient becomes dominant.

\subsection{Gas profiles for runs with constant flux jets} 
\label{S:results_const_profile}

\begin{figure}
\includegraphics[width=8cm]{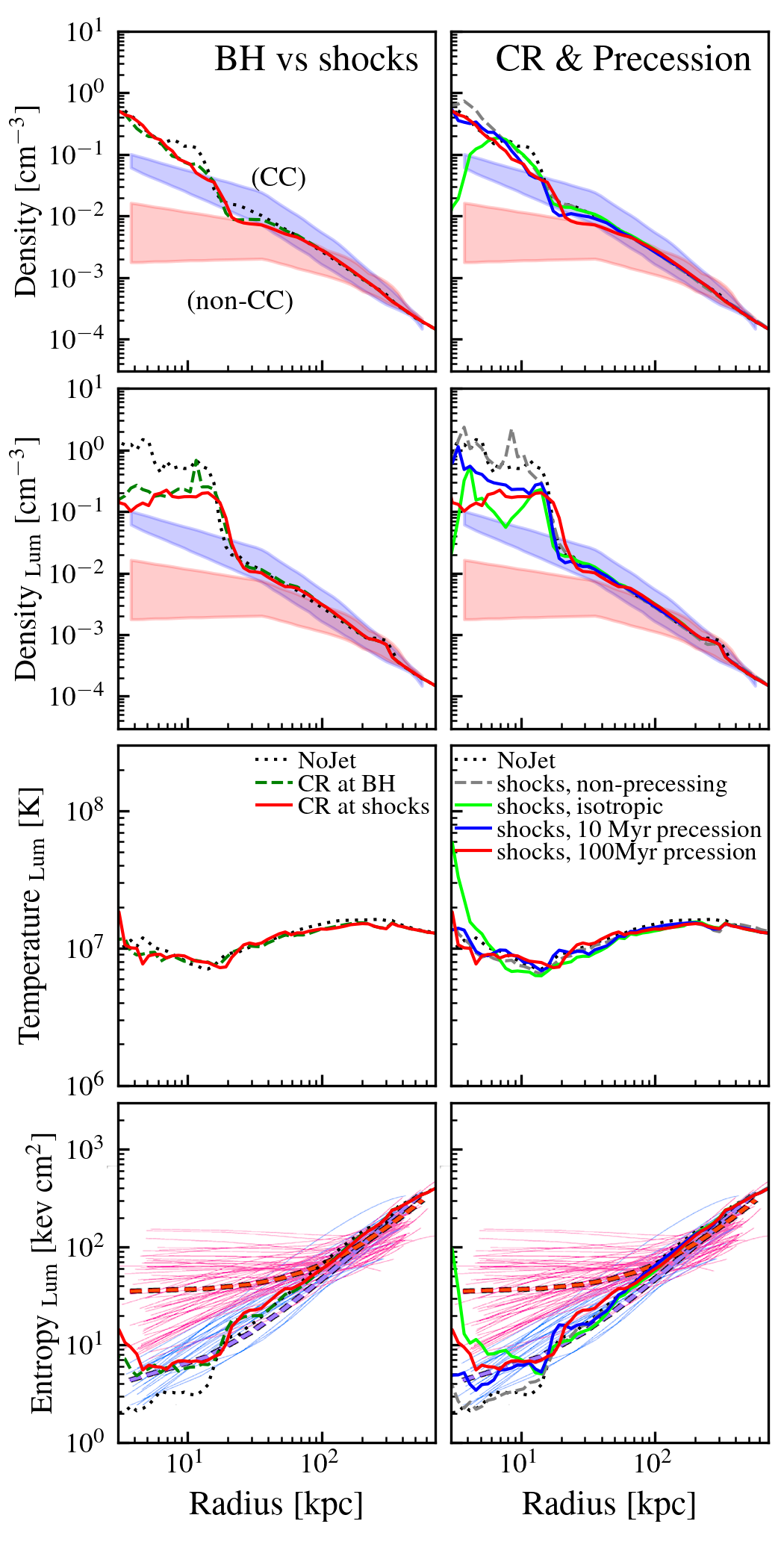}
\label{fig:profile_constant}
\caption{Mean gas density ({\em top row}), X-ray cooling luminosity-weighted density ({\em second row}), luminosity-weighted temperature ({\em third row}), and luminosity-weighted entropy ({\em bottom row}) versus radius, averaged over 0.95–1 Gyr or the last 50 Myr of the constant jet energy flux runs. The shaded regions in the first row and the light curves in the bottom row represent the observational density and entropy profiles (scaled) for cool-core (blue) and non-cool-core (red) clusters \citep{2013ApJ...774...23M}, scaled to account for halo mass differences. All runs reasonably match the observed profiles for cool-core clusters. The isotropic wind (CR-shock0.1-iso) shock heats the gas most at small radii, resulting in the hottest core.}
\end{figure}

\fref{fig:profile_constant} shows the density, temperature, and entropy profiles for all constant-flux jet runs. All profiles fall reasonably within the observed range for cool-core clusters. The non-precessing run exhibits the densest core, consistent with an unaffected cooling flow and a high star formation rate. In contrast, the isotropic wind run maintains more shocks at small radii, resulting in the most strongly heated core.


\section{Results for live-accretion runs} \label{S:results_live}

Having demonstrated the effects of different cosmic ray injection sites, jet widths, and precession rates, we now couple these to a live gravitational-torque accretion model.  When coupling the AGN jet to the accretion model, the resulting black hole accretion rate, which determines the overall jet fluxes, becomes important. 

\begin{figure*}
\includegraphics[width=16cm]{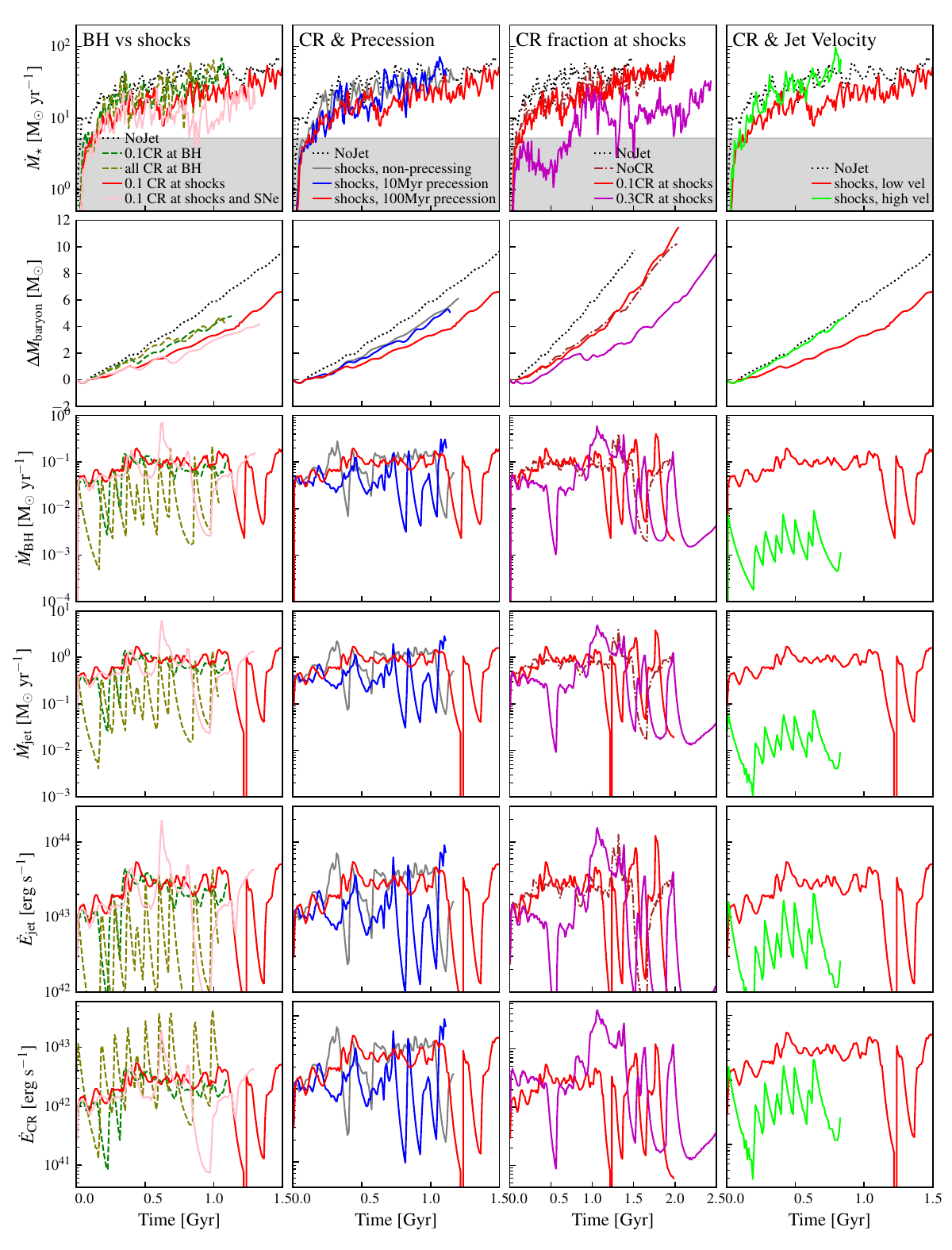}
\label{fig:sfr_bh}
\caption{Comparison of star formation rate (first row), core baryonic mass within 30 kpc (second row), black hole accretion rate (third row), jet mass flux (fourth row), jet energy flux (fifth row), and cosmic ray energy flux (bottom row) for the live-accretion runs. Unless otherwise mentioned, the plotted runs have a 100 Myr precession period and 0.1 energy in CRs.
 Injecting cosmic rays at the large-scale shock front results in a longer black hole accretion duty cycle and a higher overall cosmic ray energy flux, contributing to more effective quenching. Jets with longer precession periods push the shock front to larger radii, extending the black hole accretion duty cycle period. Increasing the cosmic ray energy fraction while keeping the jet specific energy constant results in a higher cosmic ray energy flux, leading to more effective quenching. The run with 30\% of the jet energy in cosmic rays, deposited at the shock front of a precessing jet with a 100 Myr period (CRshock0.3-pr100Myr), successfully quenches star formation with a quenching duty cycle period of $\gtrsim 100$ Myr.}

\end{figure*}

\begin{figure}
\includegraphics[width=8cm]{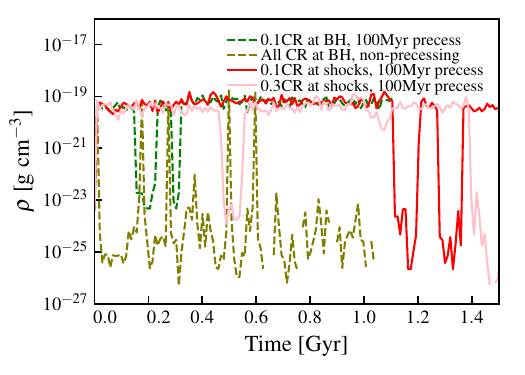}
\label{fig:density_suppression}
\caption{The density around the black hole vicinity for a subset of runs.  Injecting cosmic rays near the black hole results in episodic suppression of density in the vicinity of the black hole during the first 0.5 Gyr. When a significant fraction of the energy is in cosmic rays and deposited near the black hole, the density suppression becomes very pronounced. }
\end{figure}

\begin{figure}
\includegraphics[width=8cm]{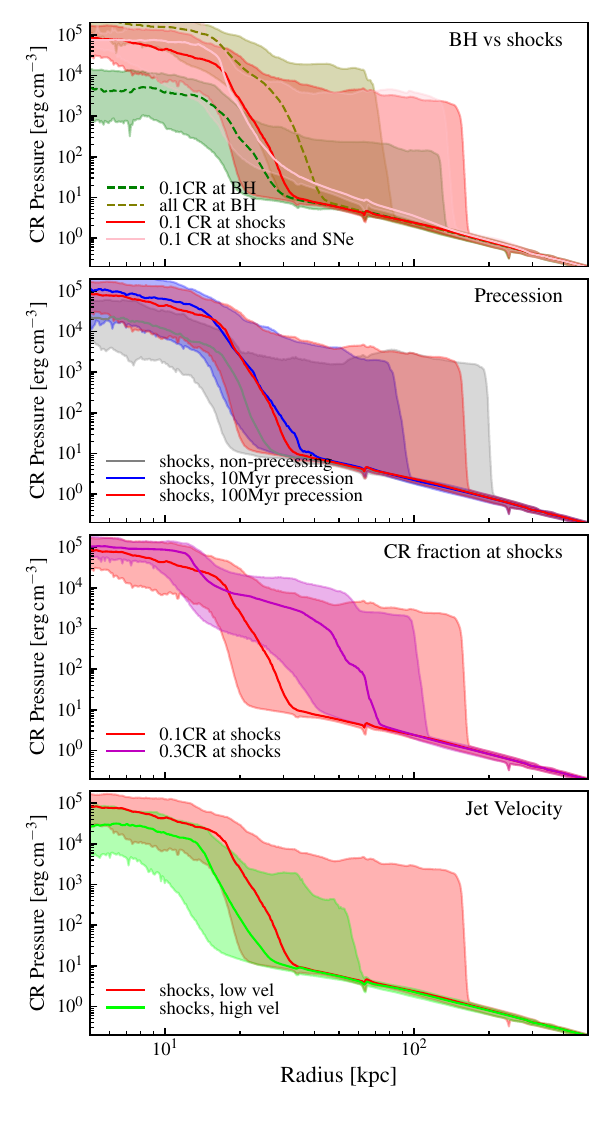}
\label{fig:cr_profile_live}
\caption{Mass-weighted median cosmic ray pressure profile (calculated over 0.95–1 Gyr or the last 50 Myr of the run) for the live accretion runs. The shaded region represents the 2\% and 98\% percentiles at each radius. Unless otherwise mentioned, the plotted runs have a 100 Myr precession period and 0.1 energy in CRs.
 Injecting cosmic rays at the large-scale jet-driven shock front (CRshock0.1-pr100Myr) produces a slightly more extended cosmic ray distribution compared to injecting cosmic rays near the black hole (CRBH0.1-pr100Myr). Faster changes in feedback direction (from fast to slow: `pr10Myr', `pr100Myr') result in more concentrated cosmic ray profiles. In the non-precessing jet case (CRshock0.1), the most extended cosmic rays are visible only in the 98th percentile, as they are concentrated in a small volume with little mass. Including cosmic rays from supernovae and increasing the fraction of AGN feedback energy in cosmic rays enhances the overall cosmic ray pressure profile. Higher jet velocity leads to a more isotropic jet cocoon, with energy concentrated at smaller radii.}
\end{figure}

\begin{figure*}
\includegraphics[width=16cm]{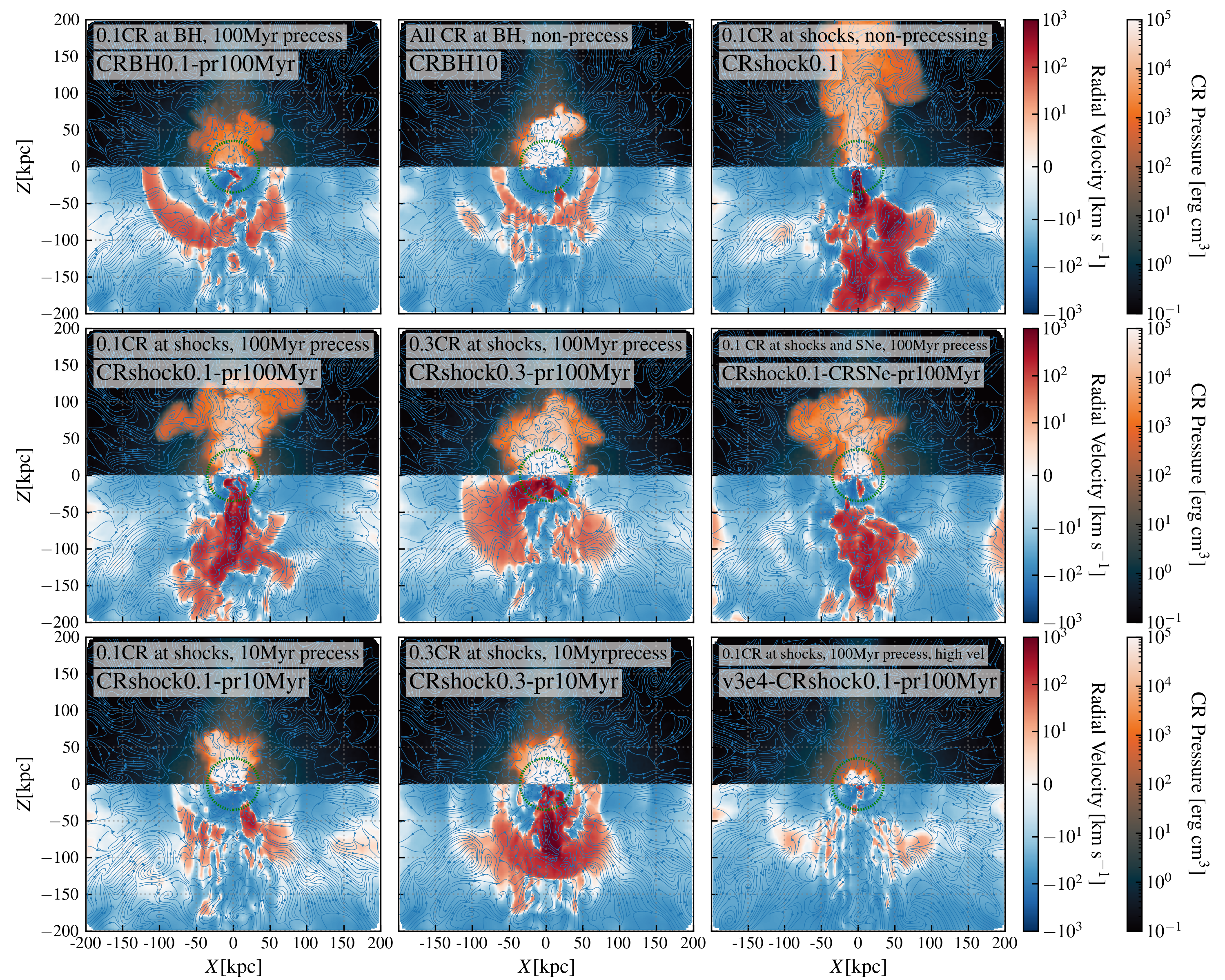}
\label{fig:morph_live}
\caption{ Morphological plot of the live accretion runs for a central slice within $|y|<10$ kpc (at 1 Gyr or at the end of the run).
 The upper half shows the cosmic ray distribution, while the lower half displays the radial velocity. Blue curves represent the magnetic field lines. Injecting cosmic rays at the large-scale jet-driven shock front (CRshock0.1-pr100Myr) results in a slightly more extended cosmic ray distribution compared to injection near the black hole (CRBH0.1-pr100Myr). Faster changes in feedback direction (from fast to slow: `pr10Myr', `pr100Myr') lead to more concentrated cosmic ray profiles. The non-precessing jet (CRshock0.1) produces the narrowest beam of cosmic rays, reaching the largest radius. Higher jet velocity leads to a more isotropic jet cocoon, with energy concentrated at smaller radii.}
\end{figure*}

\begin{figure*}
\includegraphics[width=16cm]{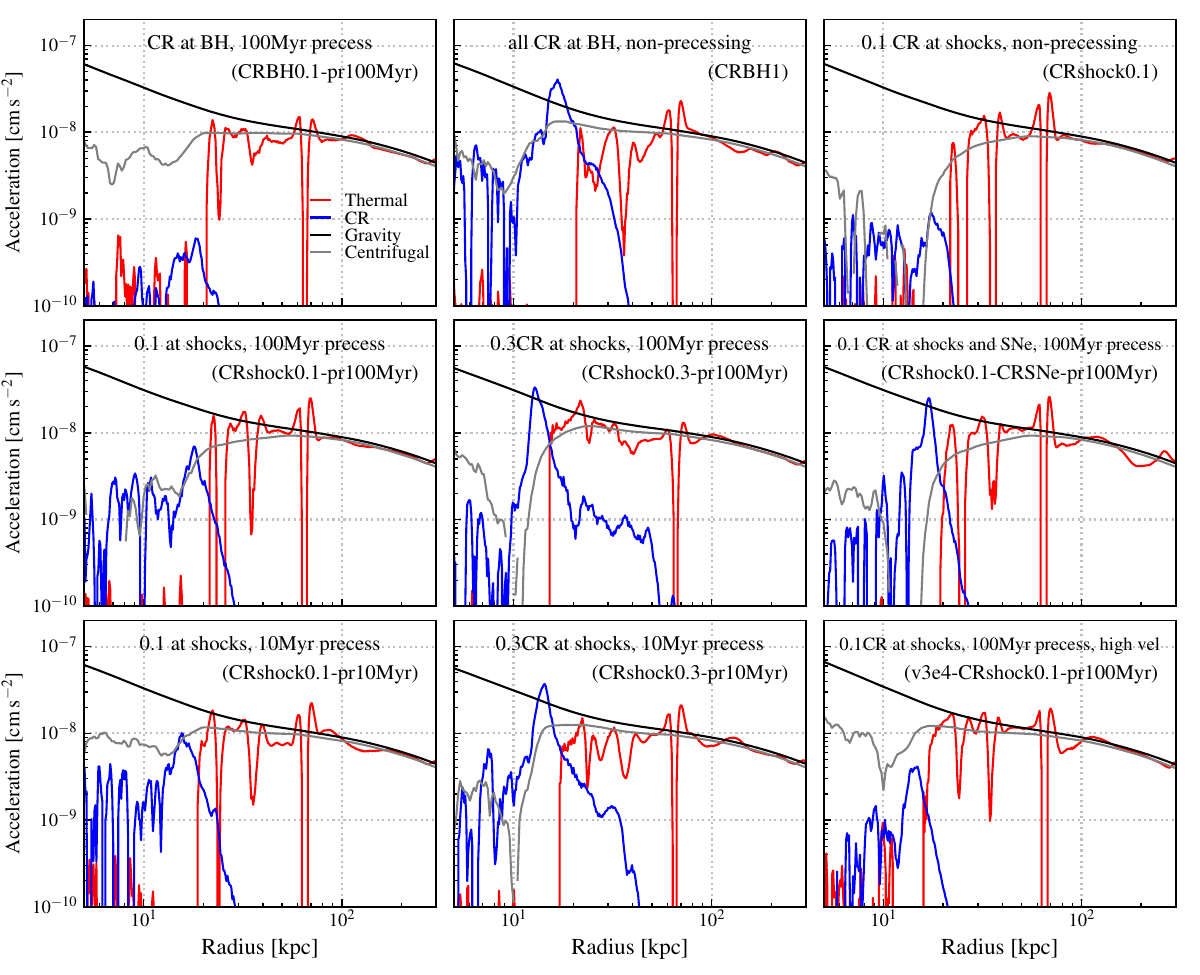}
\label{fig:cr_gradient_live}
\caption{Comparison of gravitational, rotational, thermal pressure, and CR pressure gradient acceleration in the live accretion runs (averaged over 0.95–1 Gyr or the last 50 Myr of the run). The centrifugal acceleration is defined as $GM_{\rm enc}/r^2 - v_{\rm rot}^2/r$. In the core region, where cooling is rapid, the thermal pressure gradient does not provide outward support. The stronger the suppression due to cosmic rays (e.g., from a higher CR fraction or the addition of CRs from SNe), the higher the cosmic ray pressure gradient acceleration compared to gravitational acceleration. The only exception is the run where an order-of-unity fraction of energy is deposited in cosmic rays near the black hole (CRBH1). In this case, despite the strong CR pressure gradient acceleration, the highly episodic AGN feedback results in a lower cumulative energy flux, making it less effective at quenching.}
\end{figure*}

\begin{figure*}
\includegraphics[width=16cm]{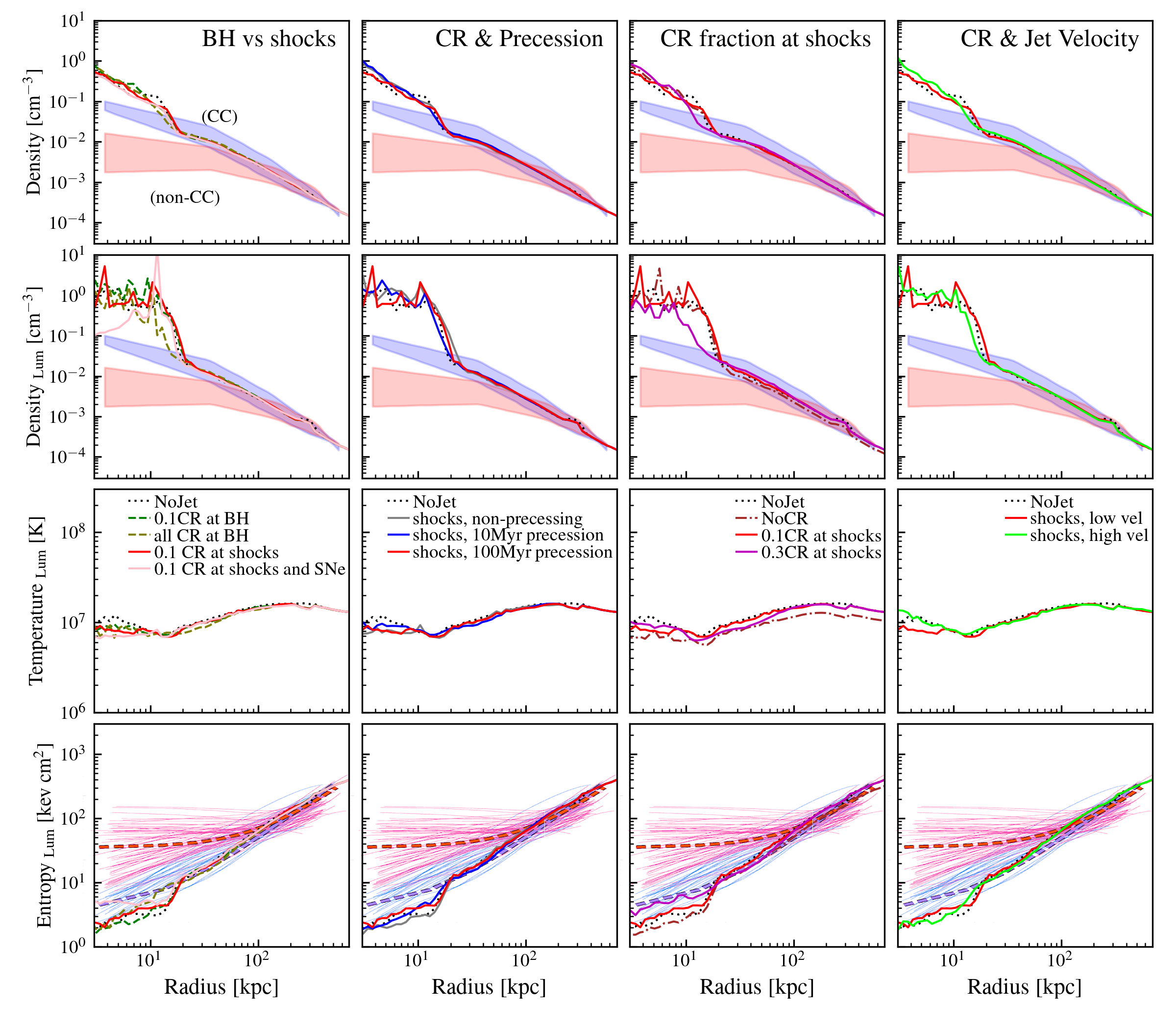}
\label{fig:profile_live}
\caption{Mean gas density ({\em top row}), X-ray cooling luminosity-weighted density ({\em second row}), luminosity-weighted temperature ({\em third row}), and luminosity-weighted entropy ({\em bottom row}) versus radius, averaged over 0.95–1 Gyr or the last 50 Myr of the  live accretion runs. Unless otherwise mentioned, the plotted runs have a 100 Myr precession period and 0.1 energy in CRs. The shaded regions in the first row and the light curves in the bottom row represent the observational density and entropy profiles (scaled) for cool-core (blue) and non-cool-core (red) clusters \citep{2013ApJ...774...23M}, scaled to account for halo mass differences. All runs reasonably match the observed profiles for cool-core clusters. A higher cosmic ray fraction deposited at the large-scale shock front results in slightly suppressed density within 30 kpc.}
\end{figure*}

In the following sections, we discuss each variation individually, including CR injection sites (\sref{sec:live_site}), jet precession (\sref{sec:live_precess}), the fraction of energy in CRs (\sref{sec:life_fraction}), and jet velocity (\sref{sec:live_vel}). For each run, we examine the black hole accretion rate, star formation rate, core baryonic mass (\fref{fig:sfr_bh}), density around the black hole (\fref{fig:density_suppression}), cosmic ray profile (\fref{fig:cr_profile_live}) and morphology (\fref{fig:morph_live}), cosmic ray pressure gradient (\fref{fig:cr_gradient_live}), and gas profiles (\fref{fig:profile_live}).


\subsection{Cosmic rays at the black hole vicinity versus the shock front}
\label{sec:live_site}

In addition to the stellar and core baryon properties, \fref{fig:sfr_bh} shows, in the 3rd to 6th rows, the black hole accretion rate, jet mass flux, total jet energy flux, and the jet energy flux in cosmic rays. The first column compares the case where cosmic rays are injected near the black hole versus at the large-scale jet-driven shock front. Injecting cosmic rays near the black hole leads to more episodic accretion, especially in the early stages of the simulation. In the run where cosmic rays carry an order-of-unity fraction of the energy and are injected near the black hole (CRBH1), the episodic deposition of cosmic rays suppresses the accretion rate by over an order of magnitude, with periods of $100-200$ Myr. A similar effect occurs within the first 0.5 Gyr even when only 10\% of the energy is in cosmic rays and is deposited near the black hole. This is primarily due to the effective suppression of density by cosmic rays, as shown in  \fref{fig:density_suppression}, which plots the density around the black hole as a function of time. Both runs exhibit several episodes of density suppression within the first 0.5 Gyr.

In contrast, in the runs where cosmic rays are injected at the large-scale shock front, this density suppression only occurs after 0.5 to 1 Gyr, resulting in a longer duty cycle period of 0.5 to 1 Gyr. This behavior is observed even when a large fraction of the energy, up to 30\%, is in cosmic rays injected at the shock front, which we will discuss later.

The varying jet fluxes, driven by different black hole accretion histories and cosmic ray coupling methods, also impact the cosmic ray distribution and star formation history. In the uppermost panel of \fref{fig:cr_profile_live},  it is clear that, with otherwise identical conditions, the run where cosmic rays are deposited at the large-scale shock front (CRshock0.1-pr100Myr) results in a more extended and overall higher cosmic ray profile compared to the run where cosmic rays are injected near the black hole (CRBH0.1-pr100Myr). This is because the former run not only has a slightly higher average energy flux but also injects cosmic rays at larger radii. This is further illustrated in the morphological plot (\fref{fig:morph_live}).  The more extended cosmic ray profile also significantly impacts the pressure gradient profile. In the former run (with injection as the shock front), the cosmic ray pressure gradient acceleration is comparable to the centrifugal acceleration, whereas, in the latter case (BH injection), the cosmic ray pressure gradient is negligible.

These differences contribute to the fact that the case with cosmic rays injected at the large-scale shock front results in slightly more suppressed star formation and less growth of the core baryonic mass, indicating more suppression of the cooling flows, as highlighted in the first two rows of \fref{fig:sfr_bh}. However, we note that according to our previous work \citep{2021MNRAS.507..175S}, a purely precessing kinetic jet without cosmic rays requires approximately $6\times10^{43}$ erg s$^{-1}$ to quench star formation, while a jet with cosmic ray energy flux of around $6\times10^{42}$ erg s$^{-1}$ is sufficient to achieve the same effect. The average jet energy flux and cosmic ray flux in neither of the runs (CRshock0.1-pr100Myr and CRBH0.1-pr100Myr) are sufficient to fully quench star formation. Note that the run with 0.3 of the energy in cosmic rays injected at the shock front appears to be sufficient and will be discussed later. Even in the run where cosmic rays carry a significant fraction of the energy flux but are injected near the black hole (CRBH1), the highly episodic suppression of black hole accretion, as mentioned earlier, results in an average cosmic ray energy flux that is still insufficient to quench star formation, despite the high cosmic ray pressure gradient (\fref{fig:cr_gradient_live}) caused by the elevated cosmic ray profile (\fref{fig:cr_profile_live}).

We emphasize that {\it cosmic rays from supernovae enhance the quenching} effect. Including cosmic rays from SNe (CRshock0.1-CRSNe-pr100Myr) results in a slightly more extended cosmic ray profile (see \fref{fig:cr_profile_live}) and noticeably greater suppression of star formation compared to the corresponding run without SNe cosmic rays (CRshock0.1-pr100Myr) (\fref{fig:sfr_bh}). At one point, the star formation rate briefly dropped to below $5 {\rm M_\odot}$ yr$^{-1}$, which we define as quiescent in this work. This is consistent with previous FIRE studies on cosmic rays from SNe \citep[e.g.,][]{chan:2018.cosmicray.fire.gammaray,hopkins:cr.mhd.fire2}.

Despite the different cosmic ray injection methods, the density, temperature, and entropy profiles remain consistent with those observed in cool-core clusters as shown in \fref{fig:profile_live}. Cosmic rays at smaller radii (from either the black hole or supernovae) slightly suppress the density profile.  When most of the jet energy is in cosmic rays, injection near the black hole can also slightly reduce the density in the 10 to 20 kpc range.

\subsection{Jet precession}\label{sec:live_precess}
The second column of  \fref{fig:sfr_bh}  compares the runs with different jet precession, with and without cosmic rays. Jet precession, combined with cosmic rays injected at the shock front, also influences black hole accretion. 
The shorter the precession period, the shorter the black hole accretion duty cycle. The 10 Myr precession run (CRshock0.1-pr10Myr) experiences an initial decline in the accretion rate within 0.3 Gyr, with a duty cycle of less than 0.3 Gyr. In contrast, the 100 Myr precession run (CRshock0.1-pr100Myr) shows the first significant decline in the accretion rate after 1 Gyr and exhibits a duty cycle ranging from 0.5 to 1 Gyr. This is consistent with the idea that cosmic rays in the faster precession jet act on gas on smaller scales, where the dynamical times are shorter. As a result, the overall duty cycle period decreases.

As shown in \fref{fig:cr_profile_live} and \fref{fig:morph_live}, changes in precession also alter the cosmic ray profiles. Consistent with our discussion in \sref{S:sfr_const}, when cosmic rays are injected at the jet-driven shock front, a longer precession periods result in more distant shocks and a more extended cosmic ray profile. This extended profile, combined with less episodic suppression of jet energy flux (due to reduced black hole accretion), leads to a slightly lower star formation rate and a smaller increase in the core baryonic mass from cooling flows, as shown in \fref{fig:sfr_bh}.

In the run with no precession (CRshock0.1), the more extended cosmic ray profile is only visible in the 98th percentile (\fref{fig:cr_profile_live}) as it occupies a small fraction of the solid angle and mass  (\fref{fig:morph_live}). As a result, as shown in \fref{fig:cr_gradient_live}, the acceleration due to the cosmic ray pressure gradient is the lowest compared to the two runs with precession. Consequently, the star formation rate (\fref{fig:sfr_bh}) is higher than in the other two runs with precession, as is the net increase in the core baryonic mass.

\subsection{Energy in cosmic rays}\label{sec:life_fraction}

We also tested different fractions of energy in cosmic rays, ranging from 0.1 to 0.3, injected at the jet-driven shock front for both cases with different precession periods (10 Myr and 100 Myr). As shown in the third column of \fref{fig:sfr_bh}, increasing the fraction of energy in cosmic rays from 0.1 (`CR-shock0.1-pr10Myr' and `CR-shock0.1-pr100Myr') to 0.3 (`CR-shock0.3-pr10Myr' and `CR-shock0.3-pr100Myr') does not significantly affect the averaged black hole accretion rate or the net jet mass and energy fluxes, regardless of whether the precession period is 10 or 100 Myr because the jet specific energy is not changed by much. However, increasing the cosmic ray energy fraction slightly reduces the black hole accretion duty cycle period during the first Gyr. 

With an overall similar total energy flux, the runs with 0.3 of the energy in cosmic rays have higher energy in cosmic rays, occasionally reaching $\gtrsim 6 \times 10^{42}$ erg s$^{-1}$, which is the required flux to quench star formation, as outlined in \citep{2021MNRAS.507..175S}. Unsurprisingly, both runs with the higher cosmic ray fraction also exhibit a higher cosmic ray pressure profile (\fref{fig:cr_profile_live}), a broader cosmic ray distribution covering a wide solid angle (\fref{fig:morph_live}), and stronger acceleration due to the cosmic ray pressure gradient  (\fref{fig:cr_gradient_live}). 
All of these factors result in more star formation suppression and less core baryonic mass growth in both runs with a higher cosmic ray fraction. In fact, both runs reach their lowest star formation rates of $\lesssim 5 {\rm M_\odot}$ yr$^{-1}$, which we define as quenched.

In particular, we ran the `CR-shock0.3-pr100Myr' case for an extended period (about 2.5 Gyr). With its longer duty cycle (around 100 Myr) and higher cosmic ray energy fraction (0.3) injected at the jet-driven shock front—both factors that aid in quenching, as discussed above—this run was the most successful in quenching star formation among all the live accretion runs. We observed a quenching duty cycle of $\gtrsim 0.5$ Gyr.

\subsection{Jet velocity}\label{sec:live_vel}

Finally, we included a run with a similar fraction of energy in cosmic rays (0.1), but with an overall higher jet specific energy, manifested in the higher jet velocity reaching $3 \times 10^4$ km s$^{-1}$ (v3e4-CRshock0.1-pr100Myr). The much higher jet velocity results in a more isotropic jet cocoon (\fref{fig:morph_live}), which does not reach as far (\fref{fig:cr_profile_live}).  This is consistent with our findings in \cite{2021MNRAS.507..175S, 2024MNRAS.532.2724S} and can be explained by the toy model describing jet propagation presented in those works.
Since much of the energy is retained at smaller radii, this leads to highly episodic suppression of black hole accretion and jet energy fluxes, as shown in the fourth column of \fref{fig:sfr_bh}. On average, the energy flux is much lower than in any run with lower specific energy. Consequently, star formation and cooling flows are barely affected.

\subsection{Factors leading to more efficient quenching}

Based on the results above, we highlight the factors that make quenching more effective for jet feedback models with a portion of energy in cosmic rays and live accretion:

\begin{itemize} 
\item \textbf{Cosmic rays at the jet-driven shock front:} Injecting cosmic rays at large-scale shock fronts results in a more extended cosmic ray distribution, which does not suppress the density near the black hole as much. This also allows for a higher overall AGN feedback energy flux, contributing to more effective quenching.

\item \textbf{Optimal jet velocity of $\lesssim 10^4$ km s$^{-1}$:} Cosmic rays are most effective when deposited near the cooling radius. If the jet velocity is much higher than this, the jet becomes lighter and isotropizes at very small radii \citep[see][]{2021MNRAS.507..175S, 2024MNRAS.532.2724S}, leading to black hole accretion being suppressed by the energy concentrated near the black hole.

\item \textbf{Optimal precession period $\lesssim 100$ Myr:} A longer precession period allows the jet to stay in a given direction for longer, driving the shock front further out. We find that a precession period of $\lesssim 100$ Myr results in most of the shock manifesting at a few tens of kpc, around the cooling radius. Depositing cosmic rays there makes quenching most efficient.

\item \textbf{Higher fraction of energy in cosmic rays:} A higher fraction of energy in cosmic rays, given similar jet specific energy, results in a greater cosmic ray energy flux, which is more effective at quenching than other energy forms.

\item \textbf{Cosmic rays from SNe:} Including cosmic rays from supernovae works synergistically with AGN cosmic rays to suppress cooling flows and star formation more effectively. 

\end{itemize}

\section{Discussion}\label{s:discussion}

\subsection{Eenergy flux required for cosmic rays to quench star formation}

Following the approach outlined in \cite{2020MNRAS.491.1190S,2021MNRAS.507..175S}, input energy of any form should, at a minimum, offset the gravitational collapse of the cooling gas:
\begin{align}
\label{eq:e_min}
\dot{E}_{\rm min}&\sim\dot{M}_{\rm cool}v_{\rm ff}[R_{\rm cool}]^2\notag\\
&\sim10^{43}\, {\rm erg\,s}^{-1} \left(\frac{\dot{M}_{\rm cool}}{100\,{\rm {\rm M}_\odot\, yr}^{-1}}\right) \left(\frac{v_{\rm ff}[R_{\rm cool}]}{300\,{\rm km\, s}^{-1}}\right)^2.
\end{align}

We also determined the required cosmic ray flux needed to suppress cooling flows by building up a pressure gradient. If cosmic ray losses are negligible and the CRs become quasi-isotropic, with an effective isotropic diffusivity $\tilde{\kappa}$ (which accounts for both streaming and advection), the steady-state cosmic ray pressure can be approximated as $P_{\rm CR}(r) \sim \dot{E}_{\rm CR} / 12\pi\,\tilde{\kappa}\,r$, as shown in several studies \citep{Buts18,hopkins:cr.mhd.fire2,2020MNRAS.496.4221J,2021MNRAS.501.3640H,2021MNRAS.501.3663H,2021MNRAS.501.4184H,2023MNRAS.521.2477B}.

To compare the outward acceleration from the CR pressure gradient, $\rho^{-1}\nabla P$, to the gravitational force ($\sim v_c^{2}/r$) where $v_c$ is the circular velocity, CR pressure can support the gas if the following condition is met:
\begin{align}
    \label{eqn:EdotCR} \dot{E}_{\rm CR} &\gtrsim  10^{43}\,{\rm erg\,s^{-1}}\,\notag\\
    &\left( \frac{10^{29}}{\tilde{\kappa}}\right)\,\left( \frac{n_{\rm gas}}{0.01\,{\rm cm^{-3}}}\right)\,\left(\frac{r_{\rm cool}}{30\,{\rm kpc}}\right)\,\left(\frac{v_{c}}{500\,{\rm km\,s^{-1}}} \right)^{2}.
\end{align}
Here, the cooling radius ($r_{\rm cool}$) is approximately 30 kpc, the density within the cooling radius ($n_{\rm gas}$) is approximately 0.01 cm$^{-3}$, and the diffusivity value used is $10^{29}\,{\rm cm^{2}\,s^{-1}}$ \citep[see][]{chan:2018.cosmicray.fire.gammaray,hopkins:cr.mhd.fire2,2021MNRAS.501.4184H}. 

These calculation provides a rough estimate for the required cosmic ray energy flux to support the gas, which is consistent with our simulations (see CR energy fluxes in \fref{fig:sft_fixed} and \fref{fig:sfr_bh}).

\subsection{Observational implications in gamma-ray wavelengths}

Given the cosmic ray energy flux above, a rough estimate of the gamma-ray flux from hadronic processes is provided by \citet{2008MNRAS.384..251G}:
\begin{align}
\dot{E}_{ \gamma}&=\int 7.51\times 10^{-16} \left(\frac{5}{6}\right)\frac{P_{\rm CR}}{\gamma_{\rm CR}-1} n_H dV\notag\\
&=\int 7.51\times 10^{-16} \left(\frac{5}{6}\right) \frac{\dot{E}_{\rm CR} r}{3\tilde{\kappa}  (\gamma_{\rm CR}-1)} n_H  dr\notag\\
&\sim 2.5\times 10^{42} {\rm erg\,s}^{-1} \left(\frac{\dot{E}_{\rm CR}}{ 10^{43} {\rm erg\,s}^{-1}}\right) \left(\frac{r}{30 {\rm kpc}}\right) \left(\frac{n_H}{0.01 {\rm cm}^{-3}}\right) \notag\\
&\gtrsim  10^{42} {\rm erg\,s}^{-1} \notag\\
&\left( \frac{10^{29}}{\tilde{\kappa}}\right)\,\left( \frac{n_{\rm gas}}{0.01\,{\rm cm^{-3}}}\right)^2\,\left(\frac{r}{30\,{\rm kpc}}\right)^2\,\left(\frac{v_{c}}{500\,{\rm km\,s^{-1}}} \right)^{2},
\end{align}
where $\gamma_{\rm CR}=4/3$ is the adiabatic index for cosmic rays, and we assume that most of the gamma-ray energy flux originates from gas within the cooling radius.

\begin{figure}
\includegraphics[width=8cm]{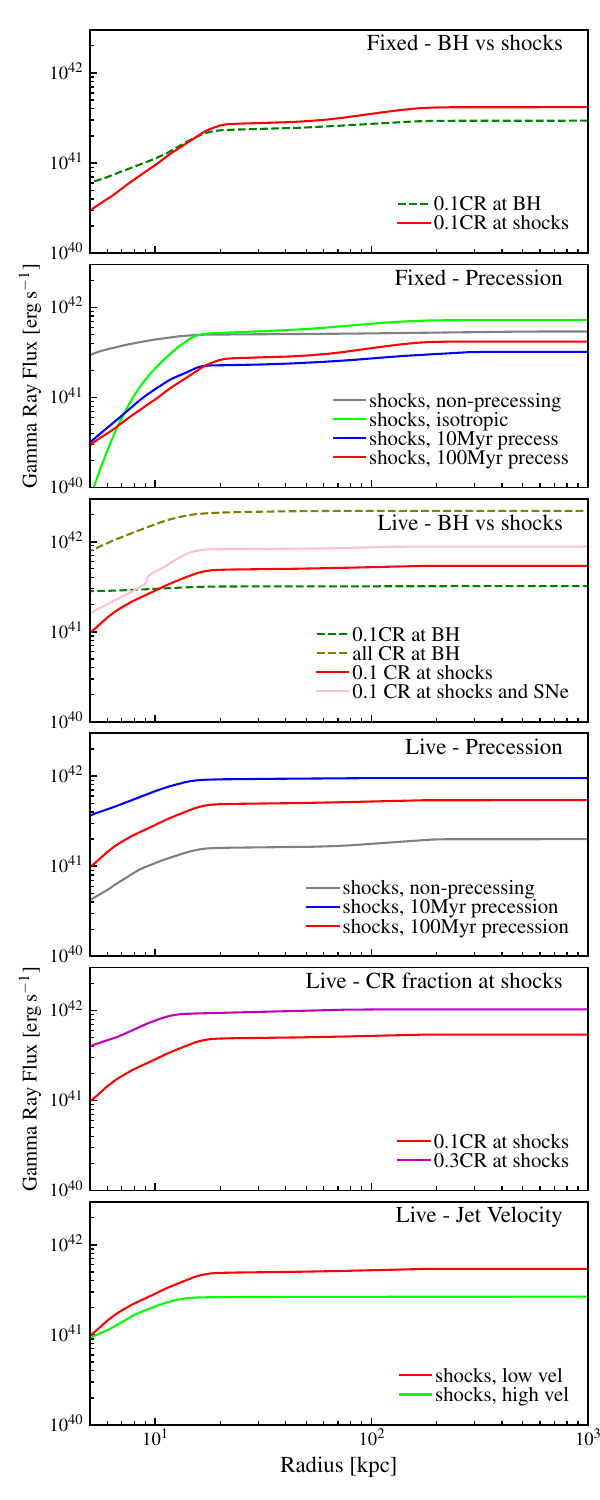}
\label{fig:gamma}
\caption{Comparison of the accumulated gamma-ray energy flux up to a certain radius for all the runs (averaged over 0.95–1 Gyr or the last 50 Myr of the run). Unless otherwise mentioned, the plotted runs have a 100 Myr precession period and 0.1 energy in CRs. All runs have gamma-ray fluxes within the Fermi-LAT limit ($\sim 1.8 \times 10^{42}$ erg s$^{-1}$, \citealt{2016ApJ...819..149A}). Decreasing the fraction of cosmic rays from unity reduces the gamma-ray flux by at least a factor of 2. Factors that result in a less cuspy cosmic ray profile, such as depositing cosmic rays at the shock front or increasing the jet precession period, lead to lower gamma-ray flux within 10 kpc. }
\end{figure}

This value is barely within the observational limits ($\sim 1.8 \times 10^{42} ,{\rm erg,s}^{-1}$) set by FERMI-LAT for the Coma cluster \citep{2016ApJ...819..149A,2019MNRAS.488..280W}.
We plot the accumulated gamma-ray flux up to a certain radius for all our runs in  \fref{fig:gamma}. Notably, although all our runs fall within this upper bound, the values from a few runs are quite close to this limit.  Some runs exhibit a much lower gamma-ray flux. 

From the plot, we can identify several factors influencing the gamma-ray flux, summarized as follows:

\begin{itemize}
\item{\bf Location of CR injection:} 
Injecting cosmic rays near the black hole results in higher cosmic ray pressure at small radii, leading to an increased gamma-ray flux in those regions, as observed in the `fixed-CRBH0.1-pr100Myr' run (first panel) and the `live-CRBH0.1-pr100Myr' run (third panel) in  \fref{fig:gamma}.

\item{\bf CR fraction in AGN feedback:} Among all the runs with constant energy flux and live accretion, the run predicting the highest gamma-ray flux is `live-CRBH1', where most of the feedback energy is in cosmic rays (third panel in \fref{fig:gamma}), even though its average cosmic ray flux from the jet is not the highest.  Reducing the cosmic ray fraction to 0.3 lowers the gamma-ray flux by at least a factor of 2, while reducing the fraction to 0.1 decreases it by at least a factor of 5.

\item{\bf Feedback geometry:} As noted in \sref{S:results_const}, isotropic winds cause most shocks to occur at small radii. Consequently, injecting cosmic rays at the shock front increases cosmic ray density and gamma-ray flux in the center within 20 kpc, as seen in the `fixed-CRshock0.1-iso' run (second panel in \fref{fig:gamma}).  At the smallest radius $<5kpc$, the suppressed density due to isotropic wind results in a lower gamma-ray flux.
In contrast, collimated feedback in the form of jets generally shifts the shocks to larger radii, reducing both cosmic ray pressure and gamma-ray flux in the center. The exception is the `CRshock0.1' run, where a strong cooling flow increases gas density within 10 kpc, thereby enhancing the gamma-ray flux.

\item{\bf Jet precession period:} A longer precession period effectively results in shocks occurring at larger radii. Therefore, when cosmic rays are deposited at the shock front, runs with a longer duty cycle period (such as the `pr100Myr' runs in the second and fourth panels of \fref{fig:gamma}) show lower cosmic ray density and reduced gamma-ray flux within 10 kpc.

\item{\bf CR from supernovae:} As expected, cosmic rays from star formation contribute to the overall cosmic ray pressure and increase the gamma-ray flux, as shown in the `live-CRshock0.1-CRSNe-pr100Myr' run (third panel in \fref{fig:gamma}).
\end{itemize}

\subsection{Limitations of our model}\label{s:caeat}

The model used in this study to inject cosmic rays at the jet-driven shock front is a simplified toy model, where we deposit cosmic rays whenever the spawned jet particles experience significant deceleration. It neglects the detailed properties of shocks and ignores shocks caused by gas that does not originate from the AGN jet. Additionally, we did not strictly conserve energy when injecting cosmic ray energy at the shock front. Although this model captures the major physical effects, as the jet particles encounter the strongest shocks in our simulation, and the injected cosmic rays contribute to less than 0.3\% of the energy budget, it remains highly simplified. A more rigorous approach, including on-the-fly shock finding and continuous cosmic ray deposition is left for future work.

In this study, we focus on testing how cosmic rays injected at the jet-driven shock front affect cooling flows, the quenching of star formation, and how they interact with other jet parameters and accretion processes. Therefore, we only included a limited variation of runs with live accretion for control experiments. We did not extensively explore variations that could significantly alter feedback efficiency, nor did we test alternative accretion models, such as Bondi accretion. While the key qualitative physical effects of launching cosmic rays at the shock front should remain robust, these variations could quantitatively impact the resulting black hole accretion history and star formation rates.

In fact, even our most quenched run only achieves star formation rates (SFR) lower than 5 ${\rm M_\odot}$ yr$^{-1}$ for short periods. When adopting a stricter definition of quenching, with specific star formation rates below $10^{-11}$ yr$^{-1}$, these periods are even shorter. Instead of claiming success for a specific parameter set, this study focuses on the effect of each physical process. Extensive parameter studies, which may optimize AGN parameters, are left for future work.

Furthermore, we only tested a single halo mass of $10^{14} {\rm M_\odot}$. As we move to higher halo masses, constraints from gamma-ray and X-ray observations could become more restrictive \citep{2024MNRAS.532.2724S}. We also leave a more detailed multi-wavelength observational constraint for future studies.

\section{Conclusions}\label{s:conclusions}

In this work, we present a toy model for injecting cosmic rays at jet-driven shock fronts using the particle spawning method. With this approach, we test the effects of large-scale shock-accelerated cosmic rays in suppressing cooling flows and quenching star formation. We also examine how injecting cosmic rays at jet-driven shock fronts interacts with black hole accretion and other jet parameters, such as velocity, opening angle, and precession. Our key findings are summarized as follows:

\begin{itemize}
\item Injecting cosmic rays at large-scale jet-driven shock fronts results in a more extended cosmic ray distribution compared to injecting them near the black hole. When coupled with black hole accretion, this leads to less suppression of the gas density near the black hole and overall higher jet energy fluxes. As a result, models with cosmic rays injected at the jet-driven shock front are more effective at suppressing cooling flows and quenching star formation.

\item For the same reason, runs with live black hole accretion and cosmic rays injected at large-scale jet-driven shock fronts result in a longer black hole accretion duty cycle compared to those with cosmic rays injected near the black hole.

\item Jet precession significantly alters the location of the shock front. The longer a collimated jet stays in a given direction, the farther the shocks are driven. A precession period of $\lesssim 100$ Myr optimally places most shocks around 30 kpc, the cooling radius of our system. Injecting cosmic rays at the shock front of jets with this optimal precession maximizes the quenching effect.

\item Increasing the jet opening angle and velocity can make the jet cocoon more isotropic and confined within a smaller radius \citep{2021MNRAS.507..175S}. Injecting cosmic rays at the shock fronts in these cases results in a more concentrated cosmic ray pressure profile.

\item For runs with live accretion, maintaining roughly the same jet-specific energy while increasing the cosmic ray fraction increases the average cosmic ray energy flux, which also leads to more effective quenching.

\item Cosmic rays from supernovae can also contribute to cosmic ray pressure and assist in quenching star formation.

\item Depositing cosmic rays at jet-driven shock fronts and having a longer precession period result in a less concentrated cosmic ray pressure profile, which also reduces gamma-ray fluxes at smaller radii.

\item As long as the cosmic ray energy fraction is $\lesssim 0.3$, the predicted overall gamma-ray flux remains well below the upper limit found by Fermi-LAT ($\sim 1.8\times 10^{42}$erg s$^{-1}$, \citealt{2016ApJ...819..149A}).

\item Our most successful run, with the most suppressed star formation, is a model where 0.3 of the jet energy flux is in cosmic rays deposited at the shock front of a precessing jet with a 100 Myr period. In this run, star formation is quenched with a duty cycle of $\gtrsim 0.5$ Gyr. 

\end{itemize}

We conclude from this work that injecting cosmic rays at large-scale shock fronts is not only physically motivated but also helps to suppress cooling flows and quench star formation when combined with jet precession. Many caveats remain to be explored in future work, along with improvements to the model (\sref{s:caeat}). 

\vspace{-0.2cm}
\acknowledgments
KS \& PN were partially supported by the Black Hole Initiative at Harvard University, which is funded by the Gordon and Betty Moore Foundation grant 8273, and the John Templeton Foundation grant 61497. The opinions expressed in this publication are those of the authors and do not necessarily reflect the views of the Moore or Templeton Foundation.  
GLB acknowledges support from the NSF (AST-2108470 and AST-2307419, ACCESS), a NASA TCAN award, and the Simons Foundation through the Learning the Universe Collaboration. 
Support for  PFH \& SP was provided by NSF Research Grants 20009234, 2108318, NASA grant 80NSSC18K0562, and a Simons Investigator Award.
RE acknowledges the support from grant numbers 21-atp21-0077, NSF AST-1816420, and HSTGO-16173.001-A as well as the Institute for Theory and Computation at the Center for Astrophysics.
YSL was supported by the NSF grant AST-2108314.
This work was performed in part at Aspen Center for Physics, which is supported by National Science Foundation grant PHY-2210452.
Numerical calculations were run on the Flatiron Institute cluster ``popeye'' and ``rusty'', Frontera with allocation AST22010, and Bridges-2 with Access allocations TG-PHY220027 \& TG-PHY220047. 
This work was carried out as part of the FIRE project and in collaboration with the LtU collaboration. LtU collaboration is supported by the Simons Foundation.

\vspace{0.3cm}
\section*{Data Availability statement}
The data supporting the plots within this article are available upon reasonable request of the corresponding author. A public version of the GIZMO code is available at \href{http://www.tapir.caltech.edu/~phopkins/Site/GIZMO.html}{\textit{http://www.tapir.caltech.edu/$\sim$phopkins/Site/GIZMO.html}}.

\bibliographystyle{mnras}
\bibliography{mybibs}

\appendix
\normalsize

\label{lastpage}

\end{document}